\documentclass[a4paper,11pt]{article}
\pdfoutput=1 
\pdfoutput=1
\usepackage{jcappub}
\usepackage{mathtools}
\usepackage{graphicx}
\usepackage{natbib}
\usepackage{multirow}
\usepackage{amsmath}
\usepackage[amssymb]{SIunits}

\usepackage{color}

\newcommand{\TM}{T_{\textrm{M}}}
\newcommand{\TCMB}{T_{\textrm{CMB}}}
\newcommand{\sigmav}{\langle\sigma_{\textrm{ann}} v\rangle}
\newcommand{\mDM}{m_{\textrm{DM}}}
\newcommand{\OmegaDM}{\Omega_{\textrm{DM}}}
\newcommand{\OmegaM}{\Omega_{\textrm{M}}}
\newcommand{\ch}{c_{\textrm{h}}}
\newcommand{\taureio}{\tau_{\textrm{reio}}}

\usepackage[T1]{fontenc}

\title{\boldmath  Dark Matter annihilations in halos and high-redshift sources of reionization of the universe}

\author[a,b]{Vivian Poulin,}
\author[a]{Pasquale D. Serpico,}
\author[b]{Julien Lesgourgues}

\affiliation[a]{LAPTh, Universit\'e Savoie Mont Blanc \& CNRS, BP 110,\\ F-74941 Annecy-le-Vieux Cedex, France.}
\affiliation[b]{Institute for Theoretical Particle Physics and Cosmology (TTK), \\ RWTH Aachen University, D-52056 Aachen, Germany.}

\emailAdd{Vivian.Poulin@lapth.cnrs.fr}
\emailAdd{Pasquale.Serpico@lapth.cnrs.fr}
\emailAdd{Julien.Lesgourgues@physik.rwth-aachen.de}

\abstract{It is well known that annihilations in the homogeneous fluid of dark matter (DM) can leave  
imprints in the cosmic microwave background (CMB) anisotropy power spectrum. However, the relevance of DM annihilations in halos for cosmological observables is still subject to debate, with previous works reaching different conclusions on this point.  Also, all previous studies used a single type of parameterization for the astrophysical reionization, and included no astrophysical source for the heating of the intergalactic medium.
In this work, we revisit these problems. 
 When standard approaches are adopted,
we find that the ionization fraction does exhibit a very particular (and potentially constraining) pattern, but the currently measurable  $\taureio$ is left almost unchanged: 
 In agreement with the most of the previous literature, 
for plausible halo models we find that the modification of the signal with respect to the one coming from annihilations in the smooth background is tiny, below cosmic variance within currently allowed parameter space.  However, if different and probably more realistic treatments of the astrophysical sources of reionization and heating are adopted, a more
pronounced effect of the DM annihilation in halos  is possible.
We thus conclude that  within currently adopted baseline models the impact of the virialised DM structures cannot be uncovered by CMB power spectra measurements,
but a larger impact is possible if peculiar models are invoked for the redshift evolution of the DM annihilation signal 
or different assumptions are made for the astrophysical contributions.
A better understanding (both theoretical and observational) of the reionization and temperature history of the universe, notably via the 21 cm signal, seems the most promising way for using halo formation as a tool in DM searches, improving over the sensitivity of current cosmological probes.}

\begin{document}
\hfill{\small LAPTH-045/15}\\
\maketitle
\flushbottom
\section{Introduction}
\label{sec:intro}

The existence of a dark component of matter in the universe, i.e. a non electromagnetically interacting form of matter, is by now well established thanks to a variety of observation in both astrophysics and cosmology:
This dark matter (DM) is necessary for instance to explain the formation of structures in the universe as we see them, and its relic density $\Omega_{\rm DM}h^2$ can be very precisely measured thanks to the cosmic microwave background (CMB) anisotropy power spectra. The quest for pinning down DM nature is however still underway, with a wide variety of techniques.

Many extensions of the standard models of particle physics (including electroweak scale supersymmetry) naturally accommodate a (quasi-)stable weakly interacting massive particle, or WIMP, that can act as excellent dark matter candidate. Additionally,  WIMP residual annihilations (or, in some models, decays) can inject sufficient ``visible'' energy that can be searched for. Notably, high energy cosmic ray and  gamma rays fluxes are routinely analysed to reveal excesses that could be attributed to a DM origin, or conservatively to constrain particle physics parameters such as the annihilation cross-section times velocity $\langle\sigma v\rangle$ (averaged over the velocity distribution, hereafter called ``the cross-section'').

Interestingly, it has been realized since the eighties that CMB observations can in fact tell us more about the nature of DM. First of all, purely gravitational arguments can lead to robust constraints on its lifetime, independently of the particle physics models (see e.g. \cite{Kaplinghat99, Ichiki04, DeLopeAmigo09, Audren14}). 
But for more specific classes of DM candidates, such as WIMPs, the CMB diagnostic power is much stronger. Annihilations (or decays, neglected from now on) inject non-thermal photons and  electrons in the intergalactic medium (IGM) that can delay the recombination and change the relic abundance of free electrons after decoupling. Hence, WIMP annihilations  can jeopardize the observed CMB temperature and polarization anisotropy angular power spectra and therefore can be constrained by an experiment like Planck. 

DM annihilations in the homogenous smooth background have been well studied and documented in the last decade \cite{Chen,Kasuya06,Zhang07,Yeung12,Padman05,Hooper09,Cirelli09,Huetsi:2009ex,Slatyer09,Natarajan10,Evoli10,Galli,Finkbeiner11,Hutsi:2011vx,Slatyer12,Giesen,Slatyer13,Lopez-Honorez:2013lcm,Slatyer15-1}. 
The most realistic calculations for WIMPs have been done by Refs.~\cite{Slatyer09,Evoli10} and updated recently in Ref.~\cite{Slatyer15-2}, where authors carefully computed how much of the initial DM particle energy is deposited into the medium, as well as how this energy is separated between ionization of hydrogen atoms, excitation of these atoms and heating of the plasma. 
They also found that the impact of DM annihilations depend sizably on the mass and the produced particles (electrons, quarks, etc.).
The Planck collaboration in a very recent paper \cite{Planck15} has reported very strong bounds on the cross-section, excluding thermal WIMPs for any standard model annihilation channel for masses up to 10 GeV, also ruling out WIMP explanations of cosmic ray lepton spectral features discussed in recent years.

Previous CMB studies mostly focused on the impact of annihilations in the averaged cosmological density field  of DM. 
However, at relatively low redshift, the DM fluid clusters under the action of gravity into virialised structures, so-called ``DM halos''.
This process increases the averaged density square $\langle \rho^2\rangle$ with respect to the square of the smooth background density, $\langle \rho\rangle^2$, while the two are nearly equal at high redshift. One could naively expect that this results in a large enhancement of the annihilation rate and therefore in a significantly bigger impact of DM annihilation on the CMB power spectra.   But the effects of halos are more subtle since, as we will see,  the way in which energy is deposited into the medium  changes as well. Thus, the modification of the bounds on DM annihilation cross-section cannot be trivially obtained.

Another interesting feature of DM halos is their possible impact on the ionization history.
In the standard picture, it is assumed that stars are the only reionization sources. Unfortunately, our knowledge of first stars formation in the universe is very rudimentary, and hence also is our knowledge of the ionization history.
The formation of halos, if it increases significantly the DM annihilation rate, could introduce a new source of reionization in the universe and leave a very peculiar imprint on the history of the ionization fraction $x_e(z)$ and temperature of the IGM $\TM$, also referred to as the matter temperature.
 In the past, this has been invoked as a way to solve a tension on the measurement of the reionization optical depth $\tau_{\rm reio}$ coming from WMAP data \cite{WMAP09} (preferring a high value of $\tau_{\rm reio}$ and therefore a relatively high ionization fraction at redshift $z>10$) and the so-called Gunn-Peterson effect as it is measured in astrophysics~\cite{Fan:2005es,Schenker:2014tda,McGreer:2014qwa} (requiring a relatively high neutral hydrogen fraction above $z\simeq6.5$ and hence pointing towards smaller values of $\tau_{\rm reio}$)~\cite{Hooper09,Cirelli09}. 
The new Planck data in the most conservative case yield $\taureio=0.079\pm0.017$ (Planck TT,TE,EE+lowP, 95\%CL~\cite{Planck15}) and therefore have reduced this tension to a $\sim$2\,$\sigma$ level, since now a single-step reionization ending at $z=6.5$ is marginally compatible with Gunn-Peterson bounds~\cite{Planck15}. Yet, it remains interesting to quantify the potential contribution of DM halos to this observable.

Only a handful of articles have investigated the impact of DM annihilation in halos,  notably~\cite{Hooper09,Huetsi:2009ex,Natarajan08,Natarajan09,Natarajan10,Cirelli09,Giesen,Lopez-Honorez:2013lcm}.  Unfortunately, previous authors follow different formalisms and are difficult to compare with each other. More importantly, they arrived at different conclusions. 
For instance, the impact of annihilation on reionization is substantial even for baseline parameters according to Ref.~\cite{Giesen}, relevant for light particles models according to~\cite{Natarajan08,Natarajan10}; while  Refs.~\cite{Hooper09},\cite{Cirelli09}, \cite{Huetsi:2009ex}, and \cite{Lopez-Honorez:2013lcm} find it to be negligible.
One of the few points on which all agree is that DM annihilation in halos cannot be the {\it only} source of reionization in the universe: even in ref.~\cite{Giesen}, an astrophysical contribution is needed at least to account for the Gunn-Peterson observations. What is less well understood is the role that DM annihilation in halos
 can play in mixed reionization scenarios, perhaps easing tensions between Gunn-Peterson observations and CMB ones. In order to clarify this situation,  we first evaluate the impact of DM annihilation in halos onto the ionization history and compute the optical depth to reionization $\tau_{\rm reio}$ in models with {\it conventional} reionization from stars: 
 We find that halos can play only a minor role, confirming the conclusions found in most previous literature. We check that Planck data do not alter the sensitivity to the halo contribution.
To clearly explain why it is so, we also characterize the impact of the annihilation in halos onto the CMB temperature and polarization power spectra, thus updating the study~\cite{Natarajan10}. In doing that, we identify and  correct a few mistakes and oversimplifying assumptions used in the previous literature~\cite{Natarajan10,Giesen}.
We also provide two major improvements over previous works: first, by adopting a phenomenological model for the star formation rate and corresponding injection
of high-energy photons, we study the dependence of the signals of interest from the astrophysical model adopted. In fact, till now all works have studied
the problem within a single type of parameterization for the astrophysical reionization history. Second, we amend an unjustified simplification in the treatment of the IGM temperature 
evolution, by adding an astrophysical source term reflecting the corresponding one in the reionization history. Finally, we discuss how and why both the ionization history and IGM temperature evolution provide more promising perpectives as DM probes via the halo term.

 Our paper is structured as follow.
In section \ref{sec:form}, we present our formalism. We review the standard recombination and reionization equations, as well as the computation of the standard energy deposition functions $f(z)$ necessary to quantify how much energy is deposited into the medium, and describe how to compute them in presence of halo formation.  Subsection~\ref{subsec:astro} is particularly original, and indulges way more than previous literature on several important aspects of the astrophysical source terms for the equations of $x_e(z)$ and $\TM(z)$.
In section \ref{sec:reio}, we compute the impact of annihilations in halos on the reionization history and revisit the question about the possible contribution of halos to solve the slight tension between CMB and Gunn-Peterson concerning the reionization optical depth $\tau_{\rm reio}$.
In section \ref{sec:cmb}, we present our results concerning the impact of DM annihilation in halos on the CMB power spectra. Section \ref{sec:disc} contains a summary of our results, as well as a discussion of possible observables where DM halos can play a non-negligible role, that would be worth studying in the future.
For the sake of a self-contained treatment, more details on the standard Peebles equations are available in appendix \ref{sec:appA}. Some remarks on the energy deposition functions and a complete comparison between the formalisms followed by different authors is developped in appendix \ref{app:fz}. Appendix \ref{sec:appC} summarizes our treatment of halo formation. Finally, appendix \ref{sec:AppendixD} contains a discussion of the reionization optical depth $\tau_{\rm reio}$ measured by Planck, compared to the {\em real} optical depth to reionization, which aims at justifying (to the best of our knowledge, for the first time) within which errors one can assume them to
coincide.

\section{Ionization and thermal evolution equations}
\label{sec:form}

Throughout this study we work with the Boltzmann code {\sf CLASS}~\footnote{http://class-code.net} \cite{Lesgourgues:2011re}, in which the two recombination codes {\sf RECFAST} \cite{Seager:1999bc} and {\sf HYREC} \cite{AliHaimoud:2010dx} are implemented.
 {\sf RECFAST} uses fitting functions to take into account the effect of highy-excited states. To deal with DM halos, {\sf RECFAST} cannot be used without extensive modifications, for its fitting functions would need to be extrapolated beyond their range of validity.
We therefore choose to use only {\sf HYREC}, in which those effects are computed at a more fundamental level without requiring interpolation.
While accounting  for highy-excited state corrections,  {\sf HYREC} (like {\sf RECFAST}) still contains a system of coupled differential equations with the same basic structure as in Peebles {\it case B recombination} model. For the evolution of the hydrogen ionization fraction\footnote{The code also follows the Helium ionization fraction, but like in most of the literature we neglect here the impact of DM annihilation on Helium recombination: it has been checked explicitely in \cite{Slatyer13,Giesen} that this effect is negligible.} $x_e$, the code integrates
\begin{equation}\label{eq:x_e}
\frac{dx_{e}(z)}{dz}=\frac{1}{(1+z)\,H(z)}(R(z)-I(z))~,
\end{equation}
where $R$ and $I$ are the recombination and ionization rates, and for the evolution of the matter temperature,
\begin{equation}\label{eq:TM}
\frac{d\TM}{dz}  =  \frac{1}{1+z}\bigg[2\TM+\gamma(\TM-\TCMB)\bigg]~.
\end{equation}
Details on these terms and coefficients are given in Appendix~\ref{sec:appA}.\\

  Energetic particles injected by any type of sources  will have three effects on the cosmic gas: direct ionization, collisional excitation (followed by photoionization by CMB photons), and heating.
These effects are taken into account by adding two terms to equation~(\ref{eq:x_e}) and one term to (\ref{eq:TM}).
The rate of the first two effects, namely direct ionization $I_{Xi}$ and excitation+ionization $I_{X{\alpha}}$, are given by:
\begin{equation}\label{eq:IonizationRate}
I_{Xi} = -\frac{\chi_i(z)}{n_H(z)E_i}\frac{dE}{dVdt}\bigg |_{\textrm{dep}}~, \quad I_{X{\alpha}} =  -\frac{(1-C)\chi_{\alpha}(z)}{n_H(z)E_{\alpha}}\frac{dE}{dVdt}\bigg |_{\textrm{dep}}~,
\end{equation}
where $E_i$ and $E_\alpha$ are respectively the average ionization energy per baryon, and the Lyman-$\alpha$ energy (see Appendix~\ref{sec:appA} for the definition of the quantity $C$).
In practice, we simply define an effective ionization rate $I_X(z) = I_{Xi}(z)+I_{X\alpha}(z)$ and add it to $I(z)$ in equations~(\ref{eq:x_e}). The heating rate term $K_h$, normalized to the Hubble rate, to be added in the square brackets at the RHS of Eq.~(\ref{eq:TM}), can 
be similarly defined as:
\begin{equation}\label{eq:Kh}
K_h =-\frac{2\,\chi_h(z)}{H(z)3\,k_b n_H(z)(1+f_{He}+x_e)}\frac{dE}{dVdt}\bigg |_{\textrm{dep}}\quad.
\end{equation}

In previous equations, $\frac{dE}{dVdt} |_{\textrm{dep}}$ stands for the energy \emph{deposited} in the plasma at redshift $z$.
It is splitted according to the {\it energy repartition fractions} $\chi_j$, with the index $j=\{ i,\alpha,h\}$ denoting ionization, excitation (through Lyman-$\alpha$ transition) and heating,  respectively. In general, the factorization into a ``universal'' deposited energy factor times the coefficients $\chi$'s (only functions of redshift) is not exact, depending on the type and energy of particles responsible for the heating, excitation and ionization.  However, it is a useful approximation for a specific source term (weak scale DM, stellar sources, etc.) and we will adopt it in the following. 

\subsection{Astrophysical source terms}
\label{subsec:astro}
In absence of DM, the high-$z$ evolution of $x_e$ and $\TM$ is the result of solving Eqs.~(\ref{eq:x_e}),~(\ref{eq:TM}), without additional source terms. 
However, the resulting evolution would be clearly unphysical in the range $z\lesssim {\cal O}(10)$. At very least, we know that the low-$z$ universe is ionized and 
relatively hot. In cosmological applications, (the bulk of) this is implicitly attributed to unspecified astrophysical sources, either unaccounted for or described
by some prescription by hand. For instance, in the  paper
\cite{Giesen} a prescription similar to the default ``single-step reionization'' incorporated in the CLASS and CAMB codes is used: below some arbitrarily chosen value 
$z_{\rm reio}$, $x_e(z)$ is cut and matched continuously to a half-hyperbolic tangent centered on $z_{\rm reio}$, reaching an asymptotic value of one (if He is neglected) at $z=0$,
with a narrow width parameter ($\delta z=0.2$) to describe a fast reionization. Apart from using such a modified $x_e(z)$, no modification at all is included in the 
evolution of $\TM$, i.e. no astrophysical sources of heating are considered. Basically all previous treatments have followed a similar approach. Here we attempt for
the first time to quantify the effect of these approximations, comparing them with a more realistic treatment of the astrophysical source terms, inspired by recent literature.

It is usually accepted that a dominant source of reionisation would be given by Lyman continuum photons from UV sources in pristine star-forming galaxies.
To account for these photons, we add a source term taken from~\cite{Robertson:2013bq,Robertson15} of the form of Eq.~(\ref{eq:IonizationRate}) to the evolution equation of $x_e$: 
\begin{equation}
\frac{1}{E}\frac{dE}{dVdt}\bigg |_{\textrm{dep}} =A_*\,f_{\rm esc}\xi_{\rm ion}\rho_{\rm SFR}(z)(1+z)^3\,
\end{equation}
where $f_{\rm esc}$ is the fraction of photons produced by stellar populations that escape to ionize the IGM, $\xi_{\rm ion}$ is the Lyman continuum photon production efficiency of the stellar population and $\rho_{\rm SFR}$ is the comoving star formation rate density.
The fiducial values for these parameters are taken from Ref.~\cite{Robertson15}, namely $f_{\rm esc} = 0.2$, $\xi_{\rm ion} = 10^{53.14}$ s$^{-1}$M$_\odot^{-1}$yr.  We also use the functional form for the star formation density rate provided in Ref.~\cite{Robertson15}, 
\begin{equation}
\rho_{\rm SFR} = a_{p}\frac{(1+z)^{b_p}}{1+[(1+z)/c_p]^{d_p}}
\end{equation}
with fit parameters $a_p = 0.01376\pm0.001$~M$_\odot$yr$^{-1}$Mpc$^{-3}$, $b_p = 3.26\pm0.21$, $c_p =2.59\pm0.14$ and $d_p=5.68\pm0.19$~\cite{Robertson15}.
Rather than varying the fitted parameters within their confidence level, for simplicity we leave the overall fudge factor $A_*$ (expected to be of order one) free to match
the measured optical depth.
A consistent treatment of such a source term in an equation analogous to Eq.~(\ref{eq:IonizationRate}) requires to specify the corresponding energy repartition functions, $\chi$'s.
We follow the prescription of Refs.~\cite{Chen,Giesen} in which 
$\chi_i = \chi_{\alpha} =(1-x_e)/3$. To motivate this choice, we remind the reader that in the simple estimate provided by Shull and Van Steenberg \cite{Shull:1982zz}  approximatively 1/3 of the energy is effectively available for ionization in a neutral gas  ($x_e=0$). The adopted expression also fulfills the obvious physical criterion that no energy is available to ionise (or excite) an already fully ionised gas ($x_e=1$). A linear interpolation is used for values in between, which corresponds to the reasonable Ansatz that the rate of ionisation is  proportional to the abundance of neutral hydrogen. The equality $\chi_i=\chi_{\alpha}$ has been  checked to be approximately true by Ref.~\cite{Chen}.

Qualitatively, stellar phenomena should also contribute to the heating of the IGM. In principle, one might expect this phenomenon to be capture by a term similar
to the one introduced above, modulo a different energy repartition fraction function. However, it is sometimes argued that heating is most efficiently achieved thanks to harder photons (X-ray band), see e.g. Ref.~\cite{Pober15}.
To account for this, we introduce a source term in the brackets at the RHS of Eq.~(\ref{eq:TM}) of the form of Eq.~(\ref{eq:Kh}), with a {\it different} normalization, taken from Ref.~\cite{Pober15},
\begin{equation}
\frac{dE}{dVdt}\bigg |_{\textrm{dep}}=3.4 \times 10^{40}f_X\rho_{\rm SFR}(z)(1+z)^3\,{\rm erg\,s}^{-1}{\rm M_\odot^{-1}\,yr}.
\end{equation}
The X-ray efficiency fudge factor $f_X$, expected to be of order $\mathcal{O}(0.1)$,
is set to the benchmark value $f_X= 0.2$. Concerning the heating repartition function, for consistency with the approximation used above for $\chi_i$ and $\chi_\alpha$ we adopt $\chi_h=(1+2x_e)/3$ as suggested by~\cite{Chen,Giesen}. To our knowledge, it is the first time that a term like this is accounted for in a study of cosmological sensitivity to
DM effects.

Finally, note that similar modifications could also be done to the equations describing helium reionisation, but we leave that for future works since {\sf HYREC} cannot be used in its current form to model such subleading effects. For the present study, we keep using a phenomenological hyperbolic tangent function to describe helium reionisation.

\subsection{Dark Matter annihilation in the smooth background}
\label{subsec:smooth}
Concerning the energy repartition functions from DM, in this work we use the recent calculations of $\chi_j(z)$ by Galli et al. \cite{Slatyer13}, that improved over former estimates \cite{Shull:1982zz,Chen}. An alternative computation has been done in Evoli et al. \cite{Evoli12}, but that of Galli et al. \cite{Slatyer13} more closely resembles  the formalism adopted here.
The energy density injection rate $\frac{dE}{dVdt}|_{\textrm{inj}}$ can be readily computed as the product among the number density of pairs of DM particles $n_{\textrm{pairs}}$, the annihilation probability per time unit $P_{\textrm{ann}}$, and the released energy per annihiliation  $E_{\textrm{ann}}$:
\begin{equation}\label{eq:EnergyInjection}
\frac{dE}{dVdt}\bigg|_{\textrm{inj}}\!\!(z) = \bigg(n_{\textrm{pairs}}=\kappa \frac{n_{\textrm{DM}}}{2}\bigg)\cdot\bigg(P_{\textrm{ann}}=\langle\sigma_{\textrm{ann}} v\rangle n_{\textrm{DM}}\bigg)\cdot \bigg(E_{\textrm{ann}} = 2m_{\textrm{DM}}c^2\bigg)~.
\end{equation}
Taking only into account the smooth cosmological DM distribution, we can write this rate as
\begin{equation}\label{eq:EnergyInjectionSmooth}
\frac{dE}{dVdt}\bigg|_{\textrm{inj,smooth}}\!\!\!\!\!\!(z)
= \kappa \rho_c^2c^2\Omega_{\textrm{DM}}^2
(1+z)^6\frac{\sigmav}{m_{\textrm{DM}}}\quad .
\end{equation}
In the equations above, $\sigmav$ is the cross-section, $n_{\textrm{DM}} = \rho_c\Omega_{\textrm{DM}}(1+z)^3$ the number density of DM particles, $\rho_c = {3H_0^2}/{8\pi G}$ the critical density of the Universe today, $\Omega_{\textrm{DM}}$ the current DM abundance relative to the critical density and $m_{\textrm{DM}}$ the DM mass.
If DM is made of self-conjugated particles, such as Majorana fermions, 
one has $\kappa=1$, which is what we shall assume in the following; if DM particles and antiparticles differ (as in the case of Dirac fermions) and are equally populated, $\kappa=1/2$ since only half of the pairs that one can form (the ones made by one particle and one antiparticle) are suitable for annihilation.\\
The response of the medium to energy injection depends strongly on the cascade of particles produced by DM annihilation, and on the epoch at which the DM particles annihilate.
This response is conveniently parametrized by a dimensionless {\it efficiency function} $f(z)$ \cite{Slatyer09} such that:
\begin{equation}\label{eq:EnergyDeposition}
\frac{dE}{dVdt}\bigg|_{\textrm{dep}}(z)=f(z)\frac{dE}{dVdt}\bigg|_{\textrm{inj, smooth}}\!\!\!\!\!\!(z) \quad .
\end{equation}
The expression of $f$ can be obtained via appropriate transfer functions $T^{(\ell)}(z',z,E)$, giving the fraction of the $\ell-$particle's energy $E$ injected at $z'$ that is absorbed at $z$, as 
\begin{equation}
f(z)=
\frac{\int d\ln(1+z')\frac{(1+z')^3}{H(z')}\sum_\ell \int T^{(\ell)}(z',z,E)E\frac{dN}{dE}\big|_{\text{inj}}^{(\ell)}dE}
{\frac{(1+z)^3}{H(z)}\sum_\ell\int E\frac{dN}{dE}\big|_{\text{inj}}^{(\ell)}d E}\label{fzexpr}
\end{equation}
where the sum runs over species (in practice $\ell$ denotes either photons or electrons), and $\frac{dN}{dE}\big|_{\text{inj}}^{(\ell)}$ is the injected spectrum of each of them in a given DM model, and is independent from $z$. 
The calculation of these functions is very involved, but it has been carefully done in \cite{Slatyer09,Slatyer12}, with a study of associated systematics presented in \cite{Slatyer13}.
While we do not indulge here in technical details, it is worth stressing a few conceptual issues concerning the meaning of $f(z)$:
\begin{itemize}
\item In the literature, the assumption that the energy released by annihilations at a given redshift is absorbed at the \textit{same} redshift is referred to as the {\it on-the-spot} approximation.
In that case, the meaning of $f(z)$ is clear: it is the fraction of energy that is absorbed by the gas, either via collisional heating or atomic excitations and ionizations.
It takes into account that a part of the energy may escape for instance in the form of neutrinos,  to some extent protons~\cite{Weniger:2013hja}, or as photons which free-stream to the present day.
The $f(z)$ factor, in this approximation, depends on DM particle model but cannot exceed 1 by definition. This approximation therefore mainly consists in considering that all absorption processes are very rapid in comparison to the Hubble time, defined as $t_H(z)= c/H(z)$.
\item However, the authors of \cite{Slatyer09} have shown that this is not true for the entire redshift
range that we are considering (and strictly speaking, not even at $z\sim1000$).
Some photons that are free-streaming at some given redshift $z'$ could be absorbed at $z<z'$.
The {\it beyond-on-the-spot} treatment consists in computing the full evolution, like in Refs.~\cite{Slatyer09, Slatyer12, Slatyer13}. The result can still be cast in the form of an efficiency function $f(z)$, simply defined {\it a posteriori} as the ratio of deposited energy to injected energy {\it at the same redshift}. In  Appendix~\ref{app:fz}, we quickly summarise the principles of the calculation performed in Refs.~\cite{Slatyer09, Slatyer12, Slatyer13}, and comment on differences with the approach of \cite{Hooper09,Cirelli09,Natarajan08,Natarajan09,Natarajan10}.
\end{itemize}
Several authors have shown that the redshift-dependence of $f(z)$ is of very little relevance for CMB constraints \cite{Galli,Hutsi:2011vx,Finkbeiner11,Giesen}.
This is because the main impact of smooth DM annihilation on the CMB is to inhibit recombination, enforcing $x_e$ to freeze out near redshift $z\sim 600$ at larger values than in standard $\Lambda$CDM.
Thus, the effects of DM annihilation is usually parameterized by a single quantity $p_{\textrm{ann}}$ defined as:
\begin{equation}
p_{\textrm{ann}}\equiv f_{\textrm{eff}}\frac{\sigmav}{\mDM}\quad\,,\:\:\:{\rm where}\:f_{\textrm{eff}}\equiv f(z=600)\,.
\end{equation}
However, it is important to keep in mind that, to correctly describe how halos of DM influenced the ionization and thermal history of the Universe, one cannot captured DM effects with only one model-independent parameter.
Furthermore, in order to be able to use the same parametrization, it is necessary to recompute these $f(z)$ functions in the presence of halo formation, and this will be the main focus of the next section.

\subsection{Dark Matter annihilation in halos}
\label{subsec:DMhalos}

\begin{figure}[!h]
\centering
\includegraphics[scale=0.5]{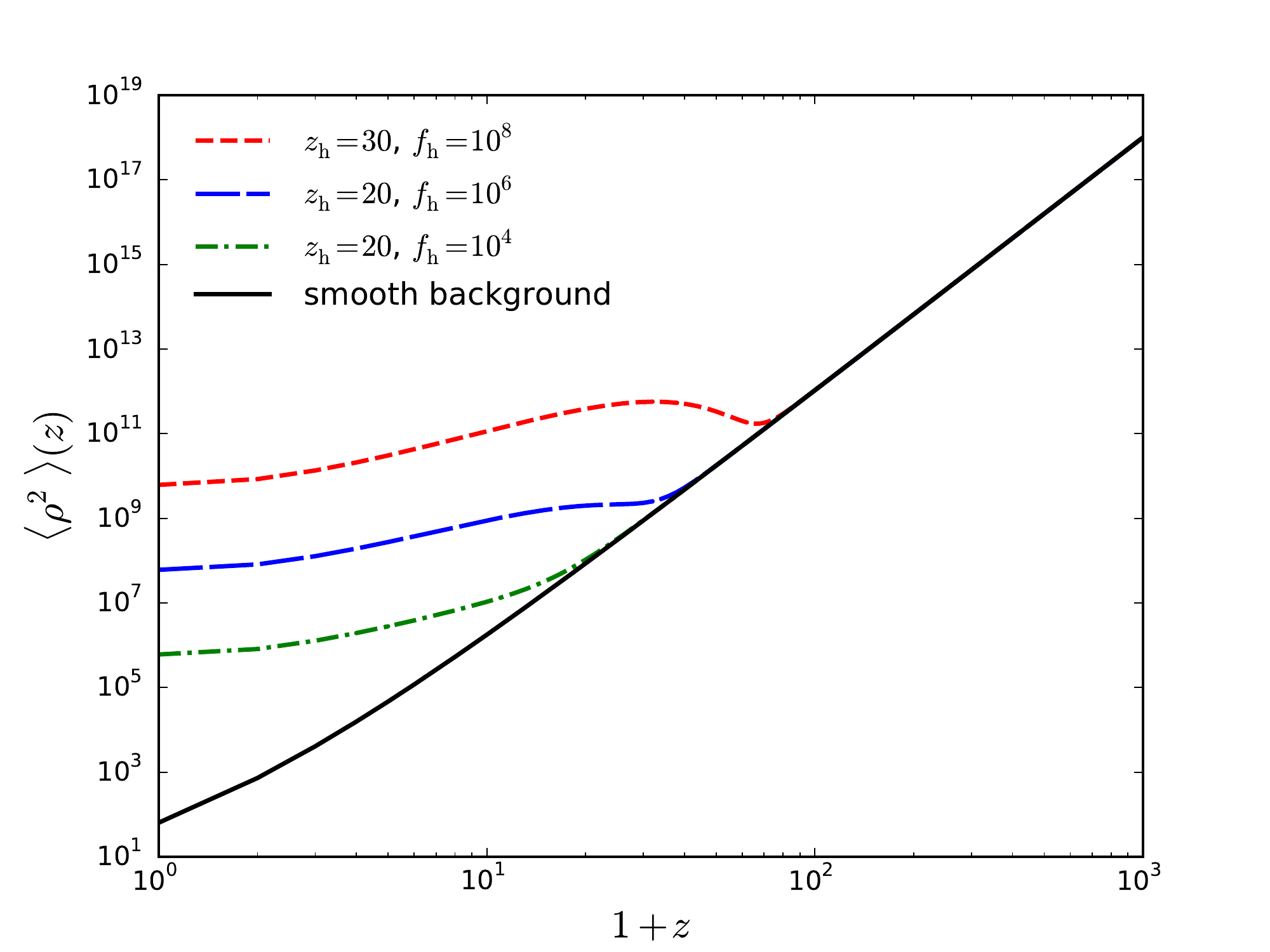}
\caption{DM squared density vs. redshift $z$ for several models of halo evolution.\label{fig:DMdensity}}
\end{figure}

The spatial average of the annihilation rate is proportional to the  average square dark matter density.
The main impact of structure formation is to increase this average with respect to the smooth background case,
by an amount usually parametrized through a boost factor $\mathcal{B}(z)$:
\begin{equation}
\langle \rho^2 \rangle (z)=  (1+\mathcal{B}(z)) \,\, \langle {\rho} \rangle^2(z).
\end{equation}
One now has:
\begin{equation}\label{eq:EnergyDepositionSmoothHalos}
\frac{dE}{dVdt}\bigg|_{\textrm{inj, smooth+halos}}=\rho^2_cc^2\Omega^2_{\textrm{DM}}(1+z)^6\frac{\sigmav}{\mDM}(1+\mathcal{B}(z))~.
\end{equation}
Several ways to compute $\mathcal{B}(z)$ have been proposed. We summarize our approach in Appendix~\ref{sec:appC}.
The two key (unknown) physical quantities are the maximal overall boost factor due to halos and the  epoch for the onset of formation of virialised objects.
The simplest choice adopted in our model was to choose as free parameters the characteristic redshift $z_h$, related to the time of halo formation (occurring near $z= 2 z_h$), and a parameter $f_h$ related to the amplitude of the boost factor today (since $\mathcal{B}(z=0)=f_h \, \mathrm{erfc}(1/(1+z_h))$, see equation (\ref{eq:BoostFactor})).
The range of values explored relies on results found in the literature, see  Appendix~\ref{sec:appC} for quantitative details.
The evolution of $\langle \rho^2 \rangle (z)$ for different values of these parameters is shown in figure~\ref{fig:DMdensity}.

Some treatments of the effect of DM (including halos) has been presented in the past: for instance, Giesen et al.~\cite{Giesen} performed this calculation on the basis of a simplified formalism developped by Natarayan~\cite{Natarajan08,Natarajan09,Natarajan10}, accounting only for energy deposition through the Inverse Compton Scattering (ICS) effect, and taking as a source for the ICS the energy injection function of the smooth case. Here however we adopt a treatment based on a more straightforward generalization of the equation~(\ref{eq:EnergyDeposition})  which is equivalent to the one reported in~\cite{Lopez-Honorez:2013lcm}. We define $f(z)$ as 
\begin{equation}\label{eq:EnergyDeposition2}
\frac{dE}{dVdt}\bigg|_{\textrm{dep, smooth+halos}}=f(z)\frac{dE}{dVdt}\bigg|_{\textrm{inj,  smooth+halos}} \quad ,
\end{equation}
where now equation~(\ref{fzexpr}) generalizes as
\begin{equation}
f(z)=
\frac{\int d\ln(1+z')\frac{(1+z')^3}{H(z')}(1+\mathcal{B}(z'))\sum_\ell\int T^{(\ell)}(z',z,E)E\frac{dN}{dE}\big|_{\text{inj}}^{(\ell)}dE}
{\frac{(1+z)^3}{H(z)}(1+\mathcal{B}(z))\sum_\ell\int E\frac{dN}{dE}\big|_{\text{inj}}^{(\ell)}}\label{fzexpr}\quad .
\end{equation}
It is clear that when setting ${\cal B}=0$ one recovers the standard expressions for the smooth contribution.

In order to assess the impact of the improved calculation of the boost factor and of the new Planck data, we computed  the energy deposition function for two baseline models with annihilation channel $\chi\chi\to e^+e^-$ and $\chi\chi\to\mu^+\mu^-$, as well as for the two  masses $\mDM = 1, 1000$ GeV, following the generalization
of the method of \cite{Slatyer12} described above and making use of the numerical tools provided by the authors\footnote{\tt http://nebel.rc.fas.harvard.edu/epsilon/}.

The generalization of the calculation for more DM models would only require to adapt the injected spectra. Since we aim at being model-independent, we selected (and limit ourselves to)  these two final state examples for two reasons: first, CMB bounds are particularly interesting for them, since light  leptonic final states are the most difficult models to constrain through other methods; second, they  represent two extreme cases for the corresponding values of $f(z)$ (high for $e$, low for $\mu$), and hence they are sufficient to bracket typical constraints, if re-expressed in terms of $\langle\sigma_{\textrm{ann}}\rangle$. 

\begin{figure}[!h]
\centering
\includegraphics[scale=0.33]{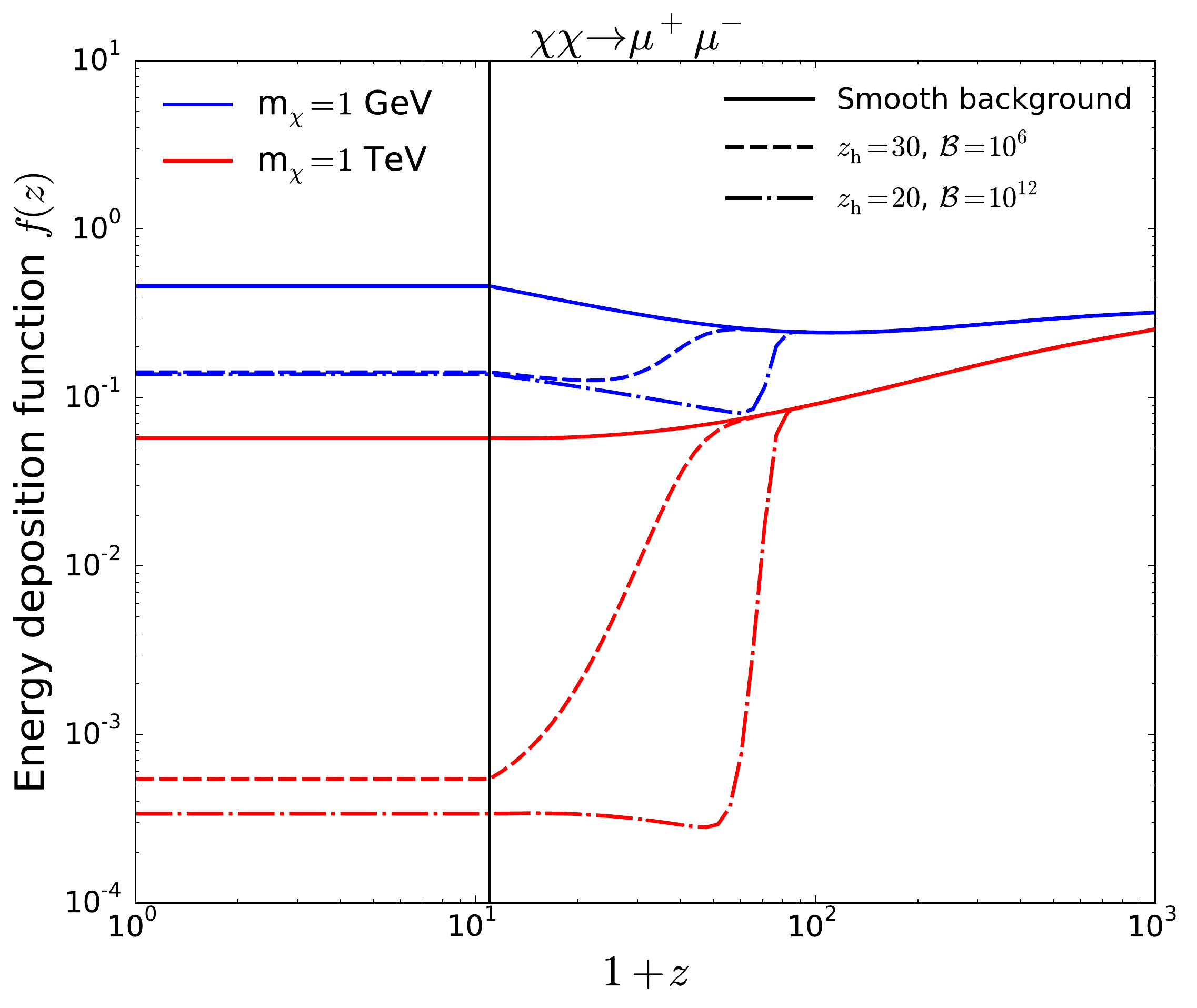}
\includegraphics[scale=0.33]{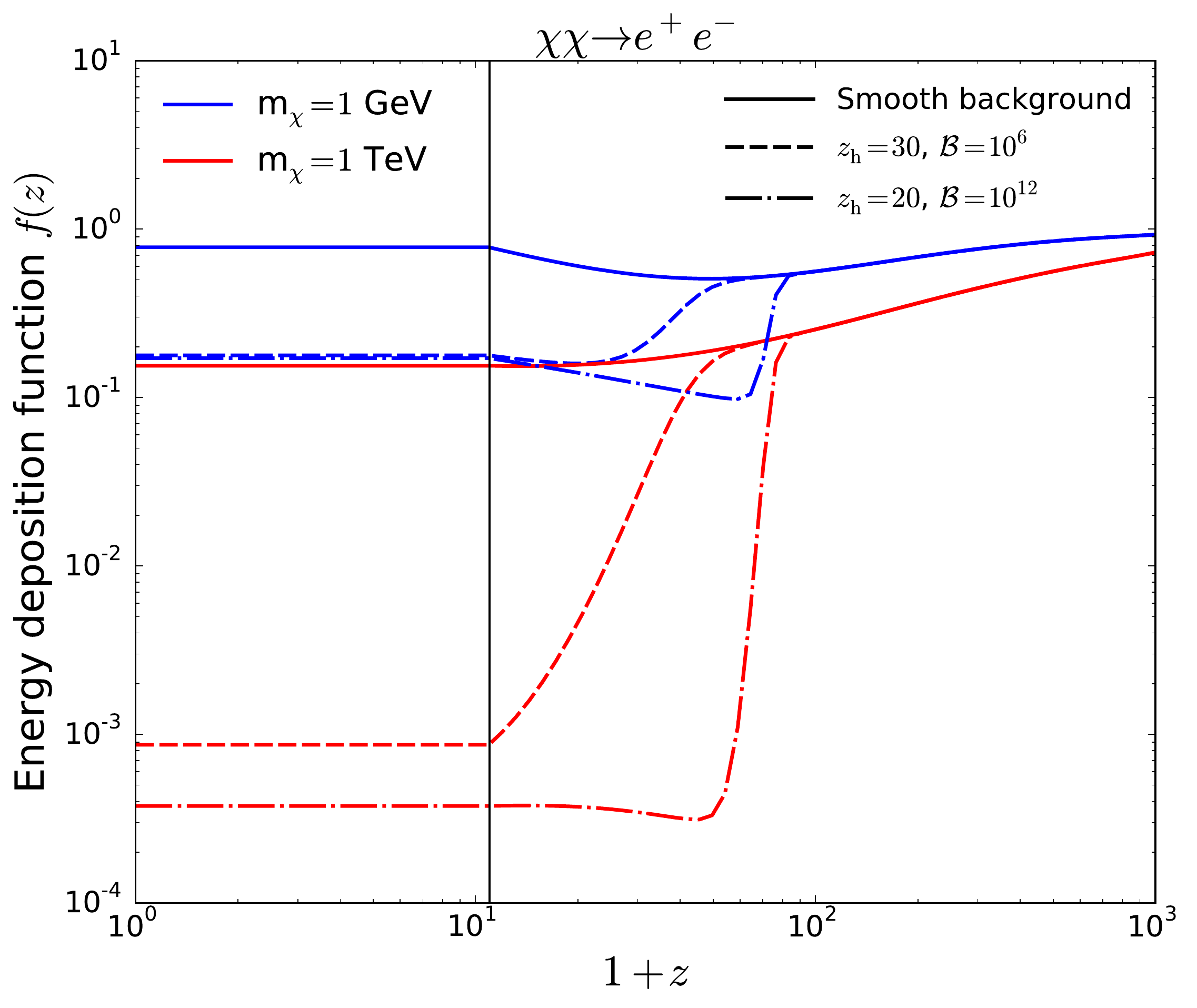}
\caption{Energy deposition functions in the two baseline models, for two values of the dark matter mass and several halo parameters, compared to the case with annihilation in the smooth background only. Below $z=10$, we assume that these functions remain constant. \label{fig:EnergyDepFunction}}
\end{figure}

Figure \ref{fig:EnergyDepFunction} shows our result for the total $f(z)$ for each baseline models, with $\mDM = 1$~GeV or $1$ TeV, and with halo parameters $[z_{\rm h} = 30, \mathcal{B} = 10^{6}  ]$ or $[z_{\rm h} = 20, \mathcal{B} = 10^{12}]$, compared to the functions computed from annihilation in the smooth background only.
Note that at high redshift, when ${\cal B} \ll 1$,  our result is asymptotically equal to that obtained in \cite{Slatyer09,Slatyer12}  in the absence of DM halos.
Note also that we only performed the calculation down to redshift $z=10$, since Ref.~\cite{Slatyer13} does not provide transfer functions below this redshift. Physically, with our definition, we do not expect big changes of the deposition function at low redshift. Hence we assume that $f(z)$ remains constant below $z=10$, as shown in Figure \ref{fig:EnergyDepFunction}. If this assumption turned out to be inaccurate, our final results would not be much affected, because observable effects at low redshift are given by the product of $f(z)$ by the factor $(1+\mathcal{B}(z))$ on which there is a huge uncertainty, and that we treat as a free parameter (see eqs .~\ref{eq:EnergyDepositionSmoothHalos} and \ref{eq:EnergyDeposition2}).

\section{Impact of high redshift sources on the reionization history}
\label{sec:reio}

\begin{figure}[!h]
\centering
\includegraphics[scale=0.23]{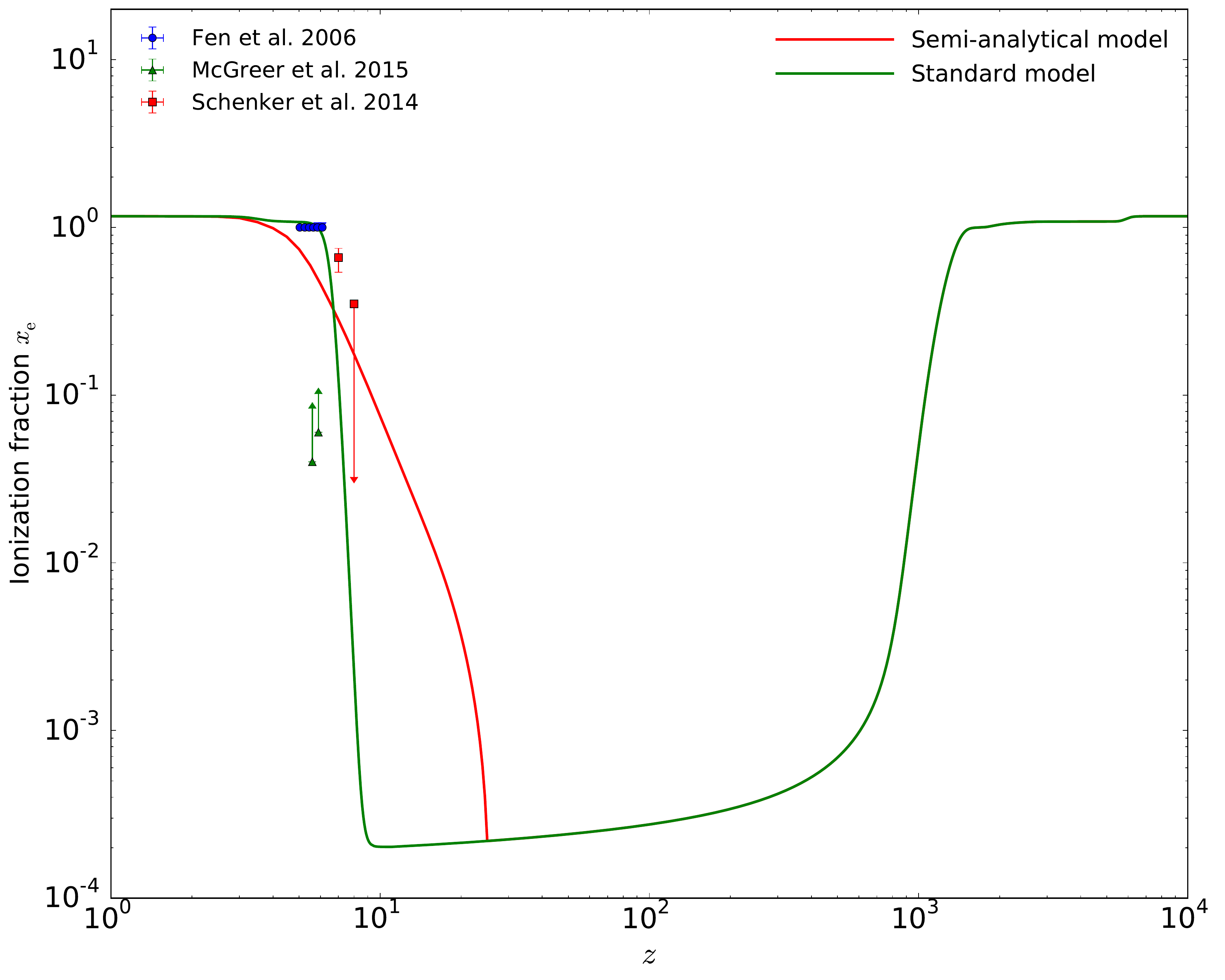}
\includegraphics[scale=0.2]{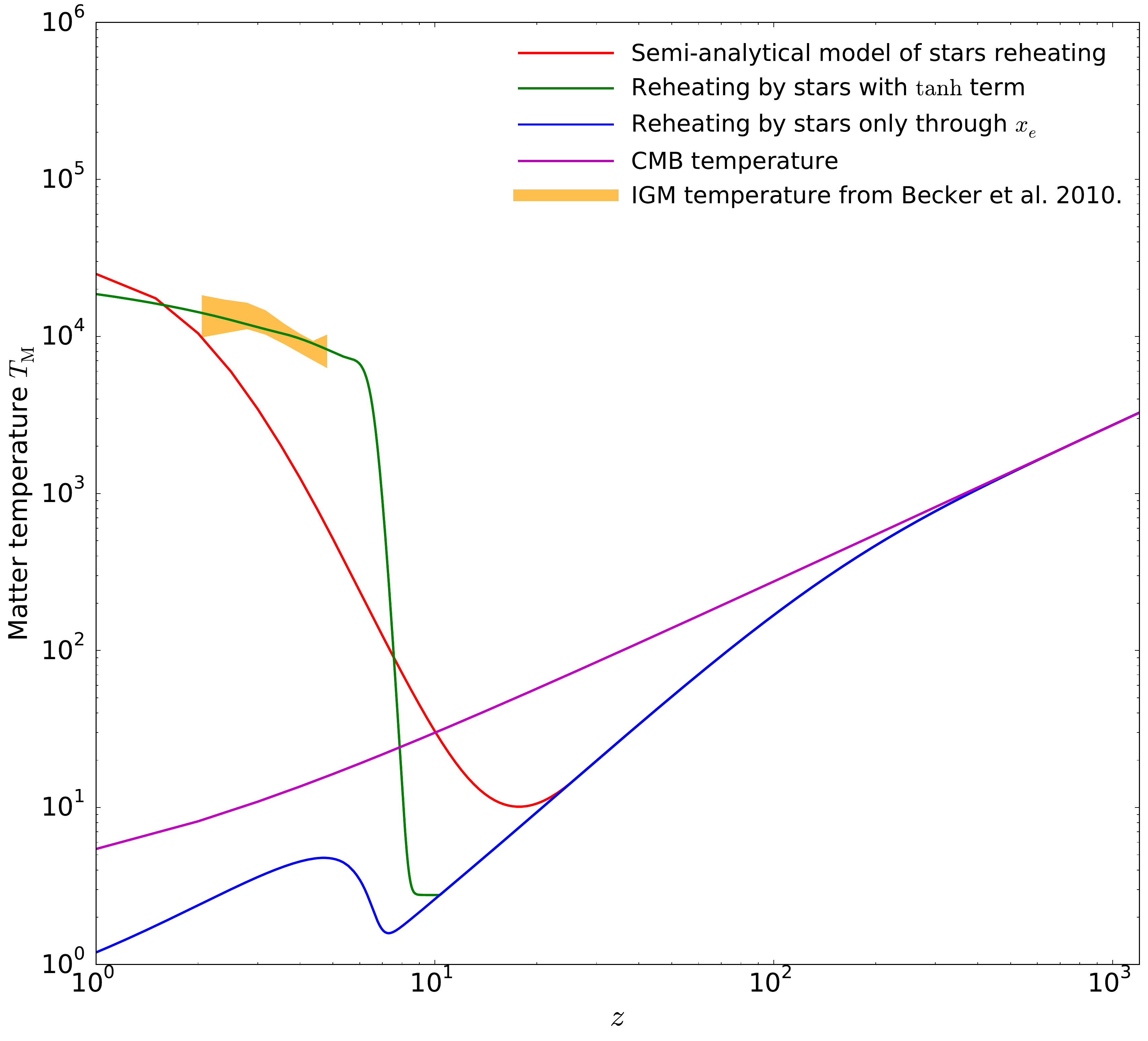}
\caption{ Evolution of $x_e(z)$ (left panel) and $\TM(z)$ (right panel) in the different approximations described in the text, for two prescriptions for describing the effect of
astrophysical sources.\label{fig:StarsxeTM}}
\end{figure}

There is no compelling reason to invoke an extra ingredient such as DM annihilation to explain the reionization history of the universe.
Actually, our knowledge of pristine star formation and its impact on the ionization history $x_e(z)$ is so rudimentary that  
currently, we can treat the ionization history $x_e(z)$ caused by star formation almost as a free function, and some room for an exotic source 
of reionization is definitely possible. To illustrate this point, in the left panel of Fig.~\ref{fig:StarsxeTM} we show two possible reionization histories of
astrophysical origin: the green curve represents the standard step-like model ``put by hand'', while the red curve represents a model inspired by actual astrophysical
data, as described in Sec.~\ref{subsec:astro}, and normalized (via the parameter $A_*\simeq 3$) so that the optical depths for the two models are the same. As far as
cosmological observations are concerned, they are essentially indistinguishable, as we will stress again in the following. The points report constraints from~\cite{Fan:2005es,Schenker:2014tda,McGreer:2014qwa}. In the right panel of Fig.~\ref{fig:StarsxeTM} we
report the corresponding gas temperature evolution, compared with the CMB temperature evolution (purple curve):  the blue curve represents the typical approximation
in which this quantity has been evolved in past literature, with only the feedback for the $x_e$ evolution accounted for (no heating source term). The green and red curves
represent the evolution of the temperature if a source term similar to the corresponding one adopted for $x_e$ is included (green: ``sudden'' heating, put by hand; red: redshift evolution
inspired by an actual astrophysical model, see Sec.~\ref{subsec:astro}). The yellow band represents some indicative constraints from ref.~\cite{Becker10}.
Our aim here is not to determine a viable heating history, rather to show the rudimentary status of these
treatments (with large uncertainties in the astrophysical term) and the large room for exotic sources of heating.

Despite the somewhat unsatisfactory situation, some consensus has been reached on important points concerning the reionization history. For instance,  in the past the question has been raised if the totality of the reionization related phenomenology
could be accounted for by DM only, but it is now acknowledged to have a negative answer. Even in Ref.~\cite{Giesen}, which finds potentially large effects at high redshift due to DM in halos, an astrophysical contribution is needed to account for the Gunn-Peterson effect, requiring the presence of a non-negligible neutral hydrogen fraction at redshift $z\sim 6.5$. On the other hand, CMB observations need the Universe to be significantly ionised at higher redshift, in order to get a correct integrated optical depth to reionization $\taureio$, compatible with measurements of the temperature and polarization spectra~\footnote{Strictly speaking,  
 the parameter called reionization optical depth by the Planck collaboration is the physical optical depth to reionization only within some assumptions (see appendix~\ref{sec:AppendixD} for details).}.  
Although early measurements (notably by WMAP 1 \cite{Spergel:2006hy}) hinted to the necessity for non-trivial reionisation history (e.g. multiple stellar populations or exotic DM contribution) because of a tension between these two observables, this is by now mostly gone: latest Planck data~\cite{Planck15} prefer a lower value of $\taureio$ and even a single-step reionization ending at $z\sim 6.5$ is marginally compatible with Gunn-Peterson bounds, reducing the tension to  a 2-$\sigma$ effect.  In a single step model, the measurement $\taureio=0.079\pm0.017$ (Planck TT,TE,EE+lowP, 95\%CL~\cite{Planck15}) translates for instance into $z_\mathrm{reio}=10.0^{+1.7}_{-1.5}$. 

However, mixed reionisation scenarios involving a relevant DM role are still of interest: for instance, one may wonder to what extent CMB upper bounds on $\taureio$ may lead to stronger constraints on DM annihilation than those coming from the smooth background. This was for instance the conclusion found in Ref.~\cite{Giesen}; even in Ref.~\cite{Natarajan10}, it was argued that at least light (few GeV's) thermal relics may have measurable effects. Articles such as~\cite{Hooper09}, \cite{Cirelli09}, \cite{Huetsi:2009ex}, \cite{Lopez-Honorez:2013lcm} find it to be instead negligible, for comparable choices of parameters.
We want to reconsider this with a state-of-the-art approach, correcting some errors and going beyond the approximations that we have identified in Refs.~\cite{Hooper09,Cirelli09,Natarajan10,Giesen} as mentioned in Sec.~\ref{subsec:smooth} and developped at the end of Appendix \ref{app:fz}. We also want to compute the full CMB power spectrum, while Refs.~\cite{Hooper09,Cirelli09} only estimated $\taureio$. This extra step may be instructive in establishing to what extent 
future data may improve over current constraints, a possibility raised for instance in Ref.~\cite{Natarajan10}.
 In the following, we fix the key parameter $p_\mathrm{ann}$ close to the current 95\%CL upper bound inferred from Planck TT,TE,EE+lowP data\footnote{We checked that the result is exactly the same with the analysis pipeline used in the Planck paper \cite{Planck15}, based on {\sf camb} \cite{camb} and {\sf cosmoMC} \cite{CosmoMC}, or using the {\sf class} version used as a baseline in this work (version 2.4.3) \cite{Lesgourgues:2011re} and {\sf MontePython} \cite{Audren12}.}~\cite{Planck15} with annihilation in the smooth background only, namely, $p_\mathrm{ann}=2.3 \times 10^{-7}$m$^3$/s/kg$=4.1 \times 10^{-28}$cm$^3$/s/[GeV/$c^2$]. If for this maximal value, we find the role of DM annihilation in halos to be negligible, then it will be {\it a fortiori} true for any viable  $p_\mathrm{ann}$ value.

 As a first sanity check that DM annihilation cannot have a dominant role in a consistent reionization history,
 having fixed $p_\mathrm{ann}$ to its maximum value, we vary the halo formation redshift $z_F$ and find the value of the boost factor ${\cal B}(z=0)$ giving a reionization optical depth of $\taureio\in [0.045,0.0113]$ (the 95\%CL  interval inferred from Planck TT,TE,EE+lowP data).
Even if the values of the parameters depends on the DM mass and annihilation channels, we find that reaching the minimal allowed $\tau_{\rm reio}\simeq 0.079$ while assuming the maximal $p_\mathrm{ann}$  requires halos to form very early and to be very concentrated, e.g., $[z_h = 50, \, {\cal B}(z=0))=10^{11}]$ for a $1000$  GeV DM annihilating into muons or $[z_h = 40,\, {\cal B}(z=0))=10^{10}]$ for a $1$ GeV DM annihilating into electrons. Cosmic ray data \cite{Fermi} and N-body simulations \cite{Serpico:2011in,Sefusatti:2014vha} are hardly compatible with ${\cal B}(z=0)\geq 10^8$. Hence, reionization from DM annihilation would require even greater halo formation redshifts than the maximal value we consider: $z_F\geq 50$. Even if very little is known about the first halos in the universe from the observational point of view, such early halo formation times do not appear realistic and are in general not considered in the literature.
We can conclude in agreement with previous studies that for conventional assumptions on annihilating DM models and on halo formation, DM annihilation cannot play a dominant role in reionizing the Universe and can at most coexist with stellar reionization.
\begin{figure}[!h]
\centering 
\includegraphics[scale=0.35]{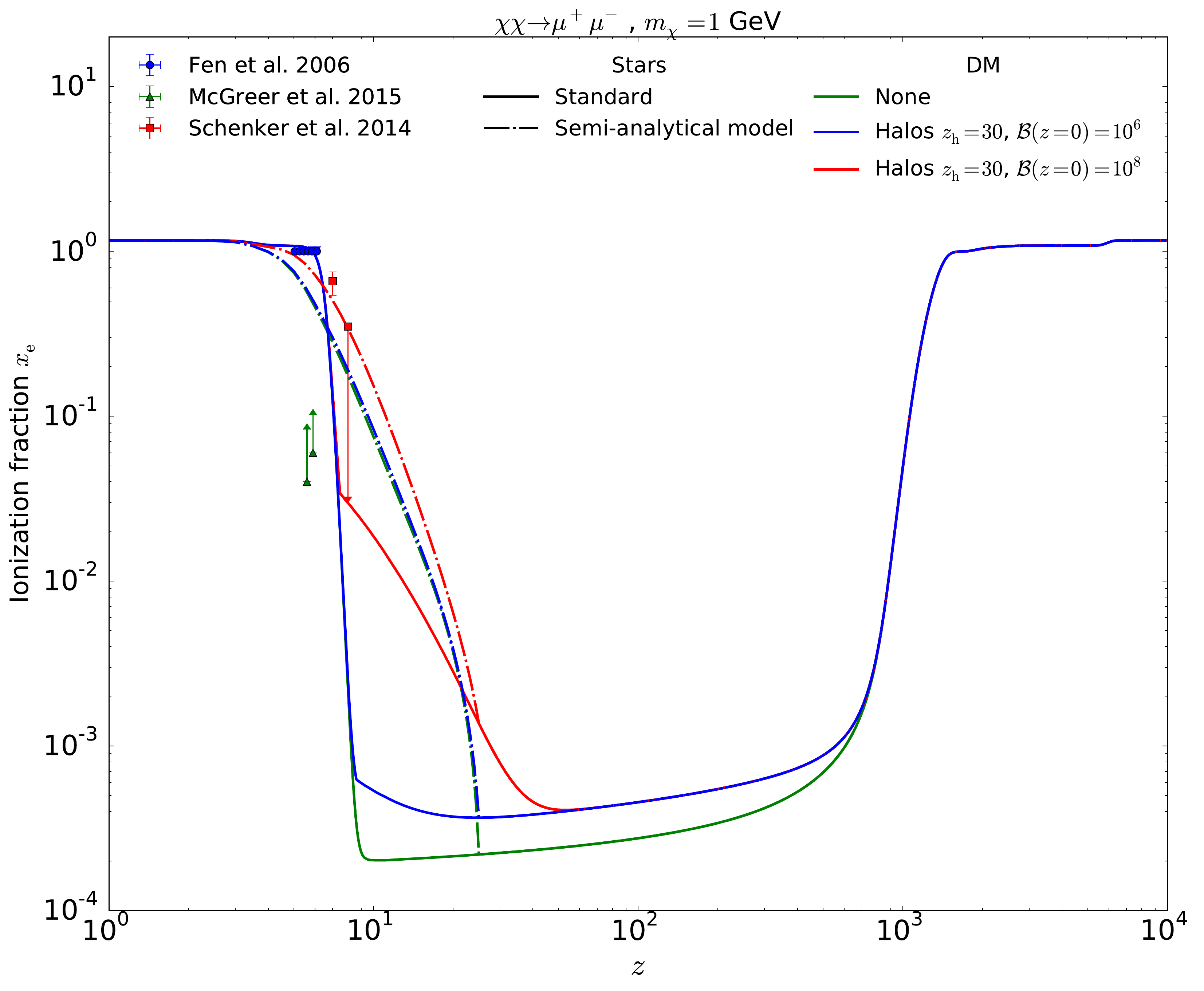}
\caption{\label{fig:xe_halo_mix} Ionization fraction $x_e(z)$ as a function of redshift for several mixed reionization models. Green lines are benchmark, purely astrophysical reionization scenarios (solid: single step; dot-dashed, phenomenological). The blue and red versions of the corresponding lines show the case where both smooth DM injection and
halo one have been added as well, with growing role of halos, respectively.
We assume $p_{\rm ann}$  fixed to its most optimistic value. }
\end{figure}

A posteriori, this justifies our choice to fix the value of $p_\mathrm{ann}$: 
If plausible reionization models involving DM annihilation could produce an exceedingly high $\taureio$, we could use the measurement of the optical depth by the Planck satellite to derive new upper bounds on this parameter. Since this is not the case, and since we have no precise information on star formation, we can always obtain the correct $\taureio$ by assuming the maximal realistic effect from DM annihilation and a complementary effect from stars. The only hope to obtain new bounds on DM annihilation from CMB observations is to analyse the full shape of the CMB temperature and polarization spectra: this will be the topic of section~\ref{sec:cmb}.

It is interesting to explore a bit further mixed reionization scenarios. In Fig.~\ref{fig:xe_halo_mix}, we show in green the  benchmark, purely astrophysical reionization scenarios (solid: single step with $z_{\rm reio}= 6.5$, as suggested by the Gunn-Peterson bound; dot-dashed, phenomenological).  The blue and red versions of the corresponding lines show the case where both smooth DM injection and halo one have been added as well, with growing role of halos, respectively. We have fixed there $p_\mathrm{ann}$ to the maximum value allowed by Planck~\cite{Planck15}. 
The free parameters are the halo ones, $z_{\textrm{h}}$ and $\mathcal{B}(0)$, besides  the DM model (annihilation channel and mass), here fixed to the muon final state and 1 GeV mass. 

\begin{figure}[!h]
\begin{center}
\includegraphics[scale=0.41]{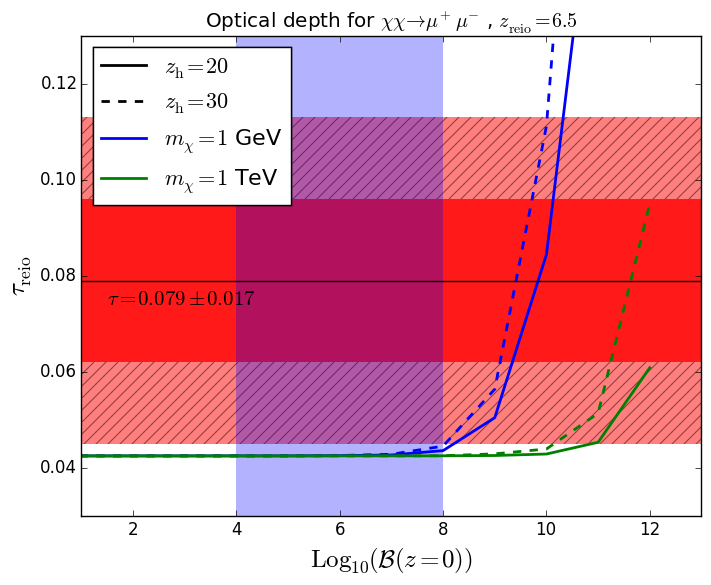}
\includegraphics[scale=0.41]{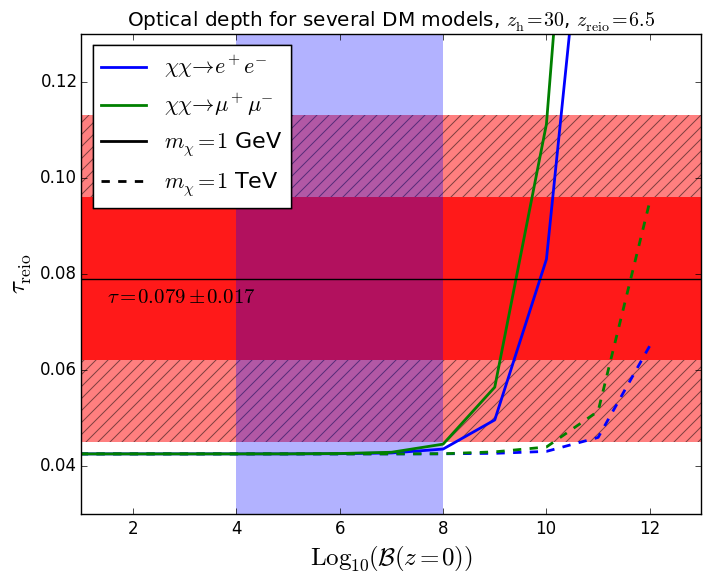}
\caption{\label{fig:tau_halo_mix} Reionization optical depth $\taureio$ in mixed reionization models, for different DM masses, annihilation channels and halo parameters.  $p_{\rm ann}$ has been fixed to its most optimistic value and the redshift od reionization from stars to $z_{\rm reio}^{\rm star} = 6.5$. The red stripe shows the most conservative bound from Planck on $\tau_{\rm reio}$ \cite{Planck15} and the blue one the most conservative interval for the values of the halo amplitudes at $z=0$ according to Refs.~\cite{Serpico:2011in,Sefusatti:2014vha}.}
\end{center}
\end{figure}

In Figure \ref{fig:tau_halo_mix}, we vary these two categories of parameters, and find $\tau_{\rm reio}$ as a function of them (in the step-like reionization scenario),
together with 68\% and 95\% confidence limits from Planck TT,TE,EE+lowP data, and bounds on ${\cal B}(z=0)$ inferred from N-body simulations by Ref.~\cite{Serpico:2011in}.
Figure \ref{fig:tau_halo_mix} shows that in absence of DM annihilation in halos, there is a marginal ($\sim 2\sigma$) tension between the model without DM annihilation in halos and the Planck bounds on  $\tau_{\rm reio}$.
At the same time, we can also see that DM annihilations in $e^+e^-$ and $\mu^+\mu^-$ for realistic values of the boost factor (blue band) do not significantly enhance $\taureio$. 
The conclusions would be similar for other annihilation channels (all being bracketed by these two)
or masses. Note that in all these models, we decided to saturate the CMB bound coming from annihilation in the smooth background. This means that we fixed the observable quantity $p_\mathrm{ann}$, but not the fundamental parameter $\langle \sigma v \rangle$. This explains why in Figure~\ref{fig:tau_halo_mix}, the effect seems to be stronger in the muon case than in the electron case; if the cross-section were fixed, the conclusion would be opposite. In this observable, the result is also independent of the reionization model adopted,
provided that they are responsible for the same optical depth. 
In all cases, our main conclusion is that one needs to push the halo contribution to the same unrealistic values as before to remove the marginal tension between CMB and the Gunn-Peterson bound.

In summary, our main conclusion is that {\it considering DM annihilation in halos formation seems neither to yield better constraints on the DM properties, nor to solve the slight tension between CMB and Gunn-Peterson data}. 
We thus essentially  confirm similar results obtained in the past, see e.g.~\cite{Lopez-Honorez:2013lcm}. Planck data are not changing these conclusions in any significant way.
On one hand, this reassures about the robustness of the reionization constraints to DM obtained by considering only the smooth contribution. On the other hand, this suggests that it is very hard to improve over them by including the relatively low-$z$ contribution from halos. Barring very different particle
physics or halo assembly histories (for some example see e.g.~\cite{Diamanti:2013bia}), it appears that the only hope to revisit this conclusion in the future and to reveal some contribution of DM halos would be to measure $x_e$ or $\TM$ as a function of $z$, especially at high redshift ($z \geq 10$), and at the same time, to improve our knowledge (both theoretically and observationally) on reionisation by the first stars.

\section{Impact of reionization histories on the CMB spectra}
\label{sec:cmb}
In this section we go one step further in the discussion of CMB sensitivity to different reionization models, beyond the simple integral constraint on $\tau_{\rm reio}$ discussed in the previous section. While unnecessary to settle the issue of current sensitivity to DM halo signals, this is useful to assess the capabilities to improve over
current constraints with future CMB data. 

 A first important point to make is that the CMB spectrum is in principle sensitive to the entire ionization history $x_e(z)$.
The main effect of DM annihilation is to delay recombination, to increase the width of the last scattering surface, and to enhance the number of residual free electrons after decoupling. 
A larger ionization fraction $x_e$ results in more Thomson scattering of photons along the line-of-sight. As long as $x_e$ remains close to its asymptotic freeze-out value, rescattering impacts CMB observables at a very small level (although not totally negligible). When stars and/or DM halo formation start, $x_e$ increases with time, and the fraction of rescattered CMB photons becomes significant. Temperature and polarization anisotropies are damped on sub-Hubble scales, and regenerated on scales comparable to the Hubble radius.

Typically, for usual models assuming single-step reionization in the range $6 \leq z \leq 12$, high $l$'s probe only the integrated parameter $\taureio$, since the temperature and polarization anisotropy spectra are mainly suppressed by a factor $e^{-2\taureio}$ (this behavior can be spoiled to some extent by a more complicated reionization history, see appendix \ref{sec:AppendixD} for details). However, small $l$'s are sensitive to the full reionization history -- especially as far as polarization is concerned.  Even in single-step models with a fixed $\taureio$, the shape of the CMB spectra in the range  $l\sim 20-40$ keeps an imprint of the details of the reionization history. In presence of DM annihilation in halos, $x_e$ tends to increase slowly at higher redshift, and a wider multipole range $l\sim 20-200$ can in principle be impacted.

\begin{figure}[!h]
\includegraphics[scale=0.33]{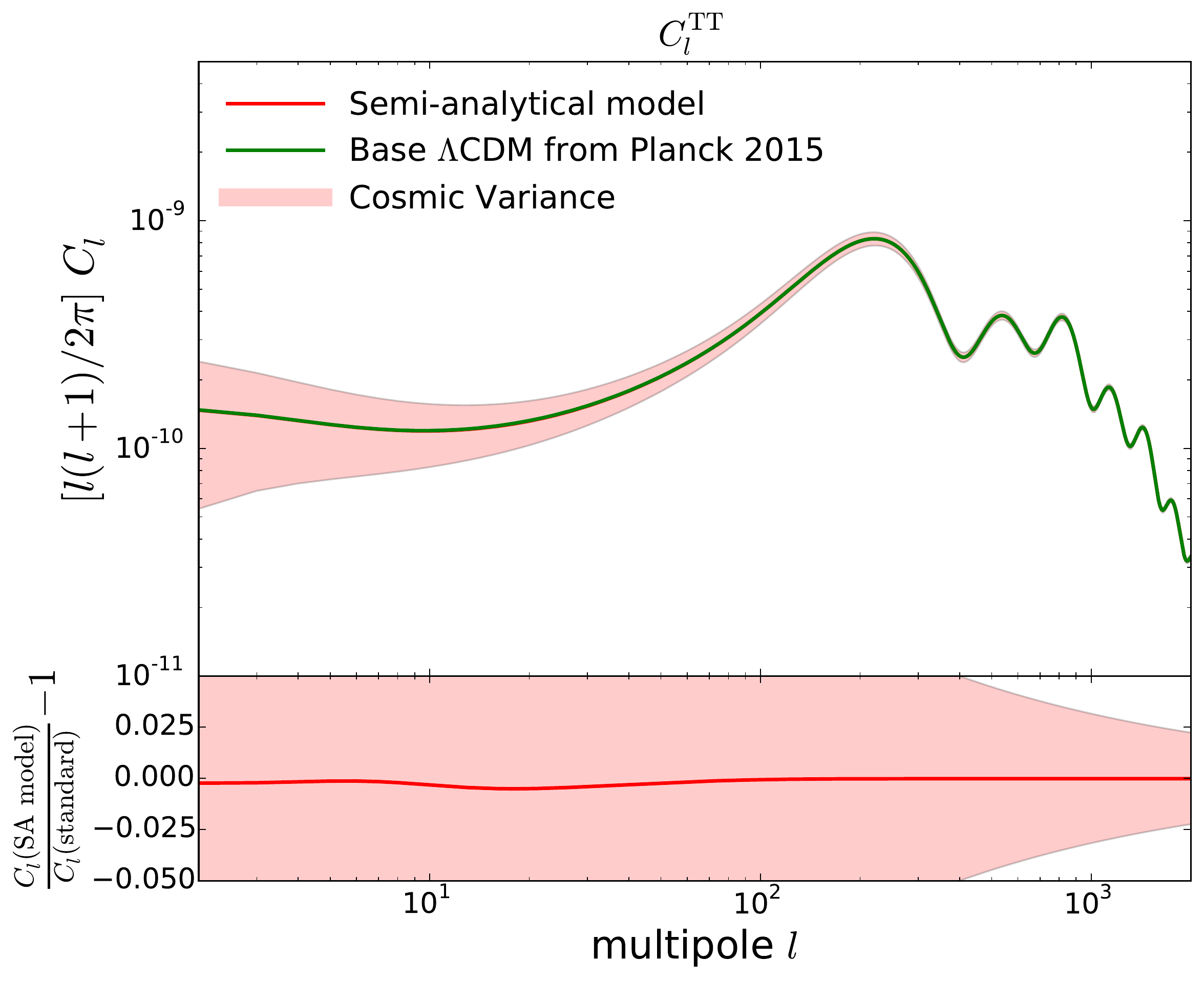}
\includegraphics[scale=0.33]{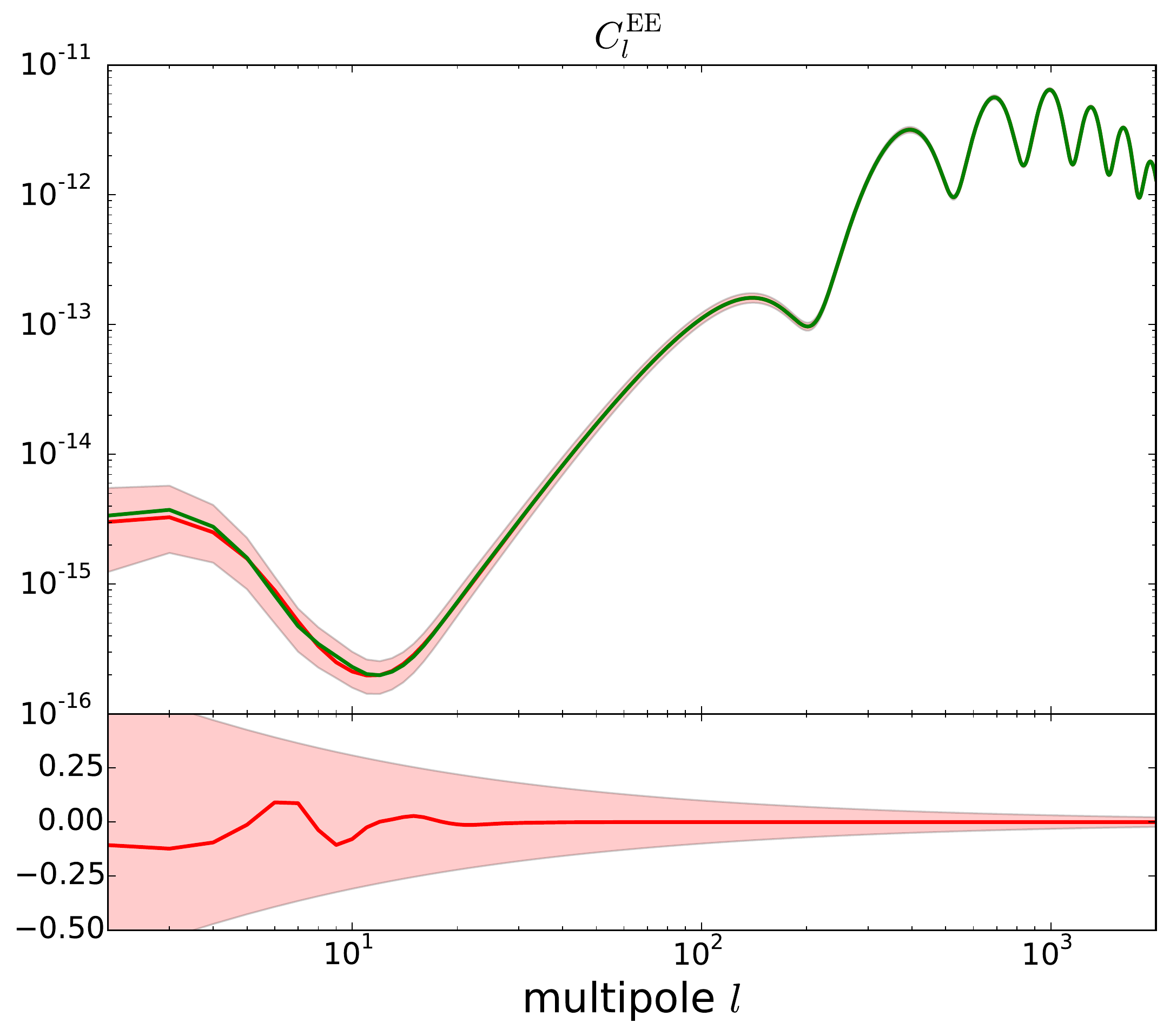}
\caption{\label{fig:clstars} Upper pannels $-~C_l^{TT}$, $C_l^{EE}$ for the two models of astrophysical reionization described in Sec.~\ref{subsec:astro}. The shaded area represents cosmic variance.
The baseline $\Lambda$CDM model is assumed, with  $\taureio = 0.079$ and $\theta_s = 1.04077\times10^{-2}$ fixed in agreement with Planck measurement (TT,TE,EE+lowP~\cite{Planck15}).
Lower pannels $-$ Relative difference between the two models.}
\end{figure}

In Fig.~\ref{fig:clstars} we show the temperature (left) and EE polarization (right) multipole spectra computed for the two models of astrophysical reionization described in Sec.~\ref{subsec:astro}, with the bottom panels showing the relative difference between the two models, compared with the cosmic variance (shaded areas). 
They have been produced with {\sc class} version 2.4.3 for a few DM models with fixed $\theta_s= 1.04077\times10^{-2}$ and $\tau_\mathrm{reio}=0.079$, in agreement with Planck measurement (TT,TE,EE+lowP~\cite{Planck15}). The same $\theta_s$ is obtained by adjusting $H_0$ (with fixed $\omega_b$, $\omega_\mathrm{cdm}$).
The fact that the variations shown in these plots are well within the limits of cosmic variance exemplifies why, despite the fact that a sensitivity to $x_e(z)$ is present in principle, it is considered to be hopeless to infer information on different astrophysical reionization scenarios via CMB observations, and why to
a large extent in this framework it is an excellent approximation to assume that CMB is only sensitive to $\taureio$. Nonetheless, note that a greater sensitivity to $x_e(z)$ of the EE polarization
with respect to TT spectrum is still manifest.

\begin{figure}[!h]
\includegraphics[scale=0.33]{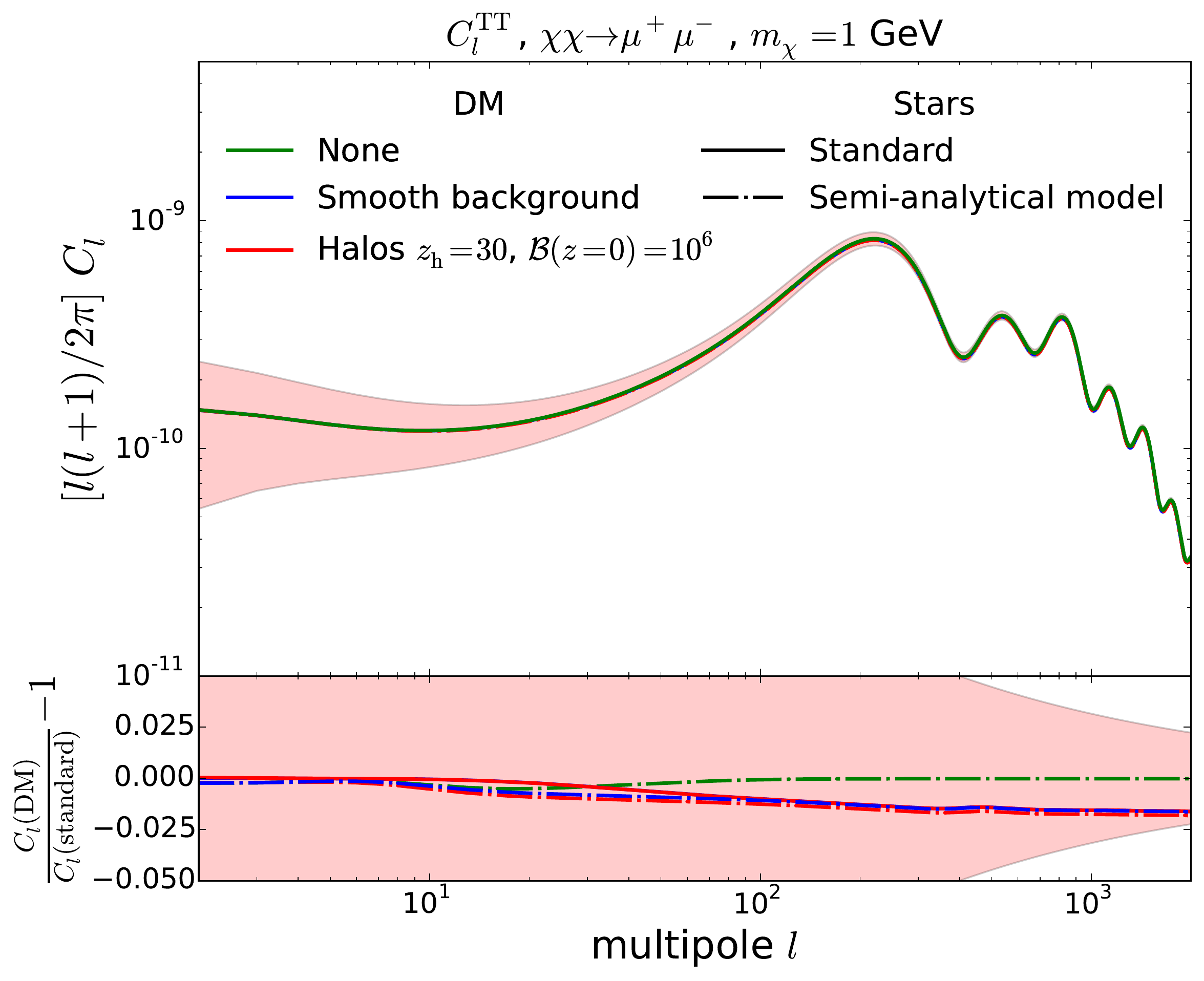}
\includegraphics[scale=0.33]{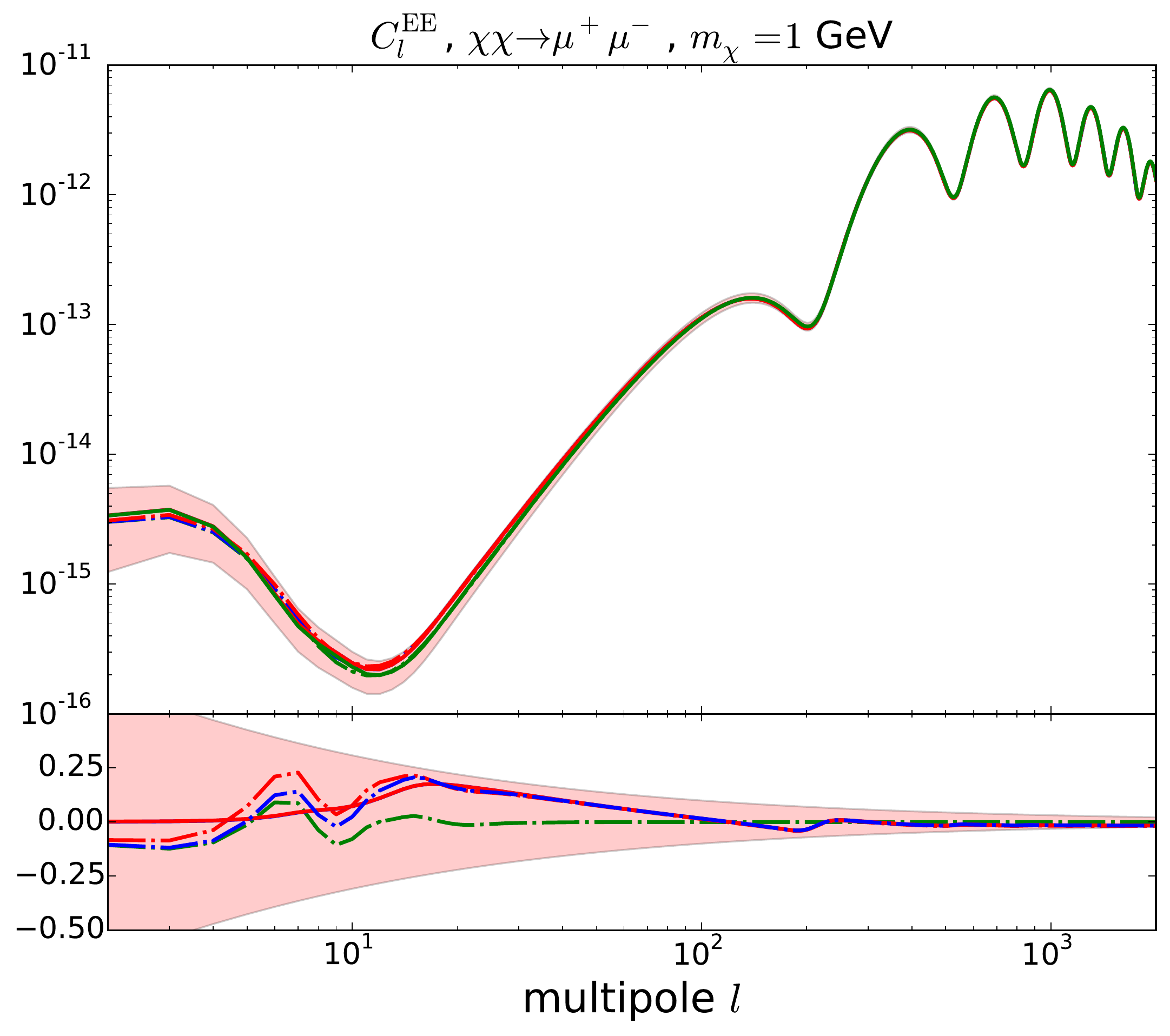}
\caption{\label{fig:cl_halos} As in Fig.~\ref{fig:clstars}, but with the DM effects (both smooth and in halos) now added, for a single reference case of halo boost factor.}
\end{figure}

\begin{figure}[!h]
\includegraphics[scale=0.33]{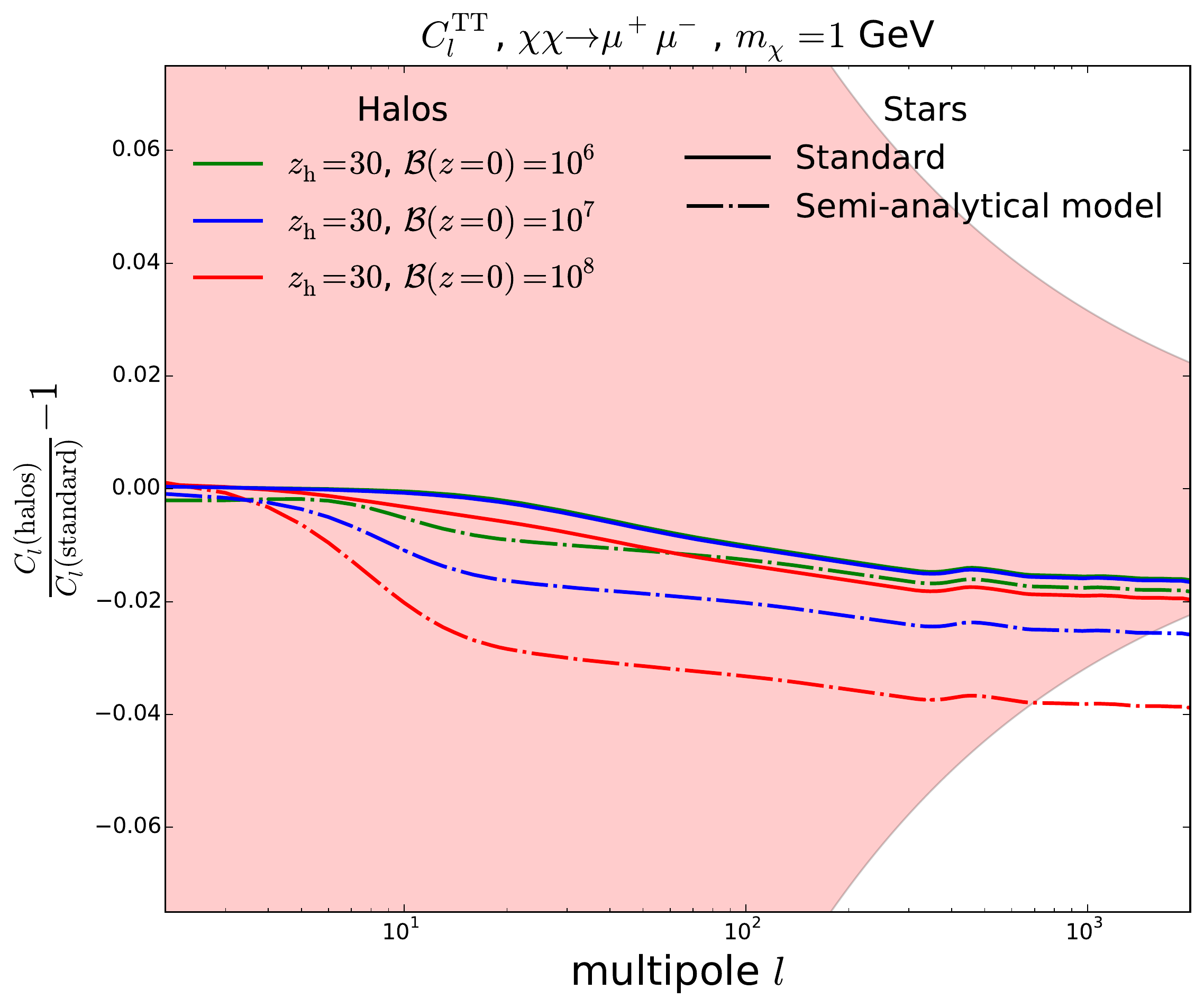}
\includegraphics[scale=0.33]{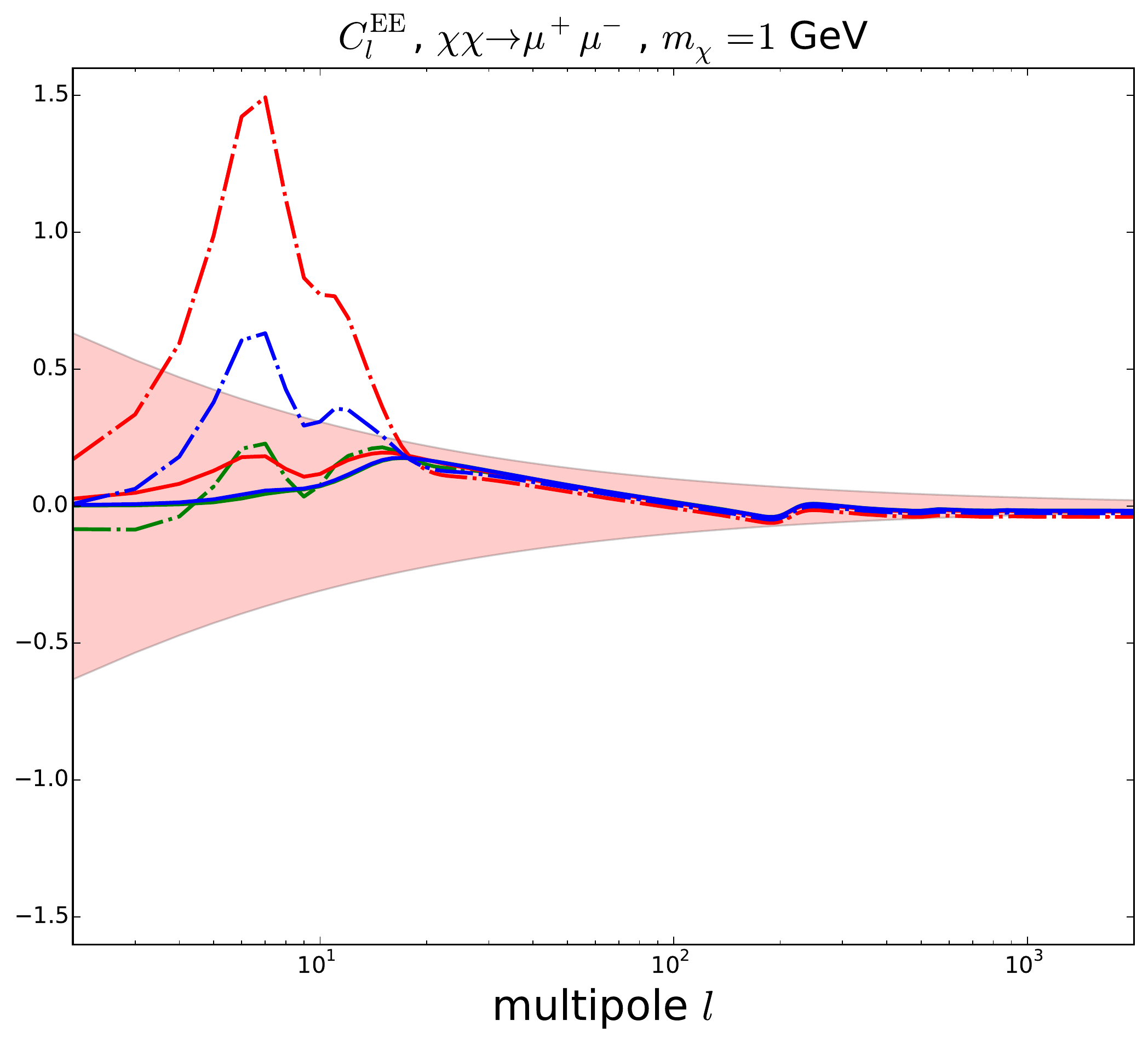}
\caption{\label{fig:cl_halos_res} Zoom on the residuals with respect to the baseline models reported in Fig.~\ref{fig:clstars}, with several assumptions for the halo contribution reported and for two choices of the astrophysical reionization.}
\end{figure}

The situation is partially altered if DM annihilation in halos is added. In Fig.~\ref{fig:cl_halos}, we show plots similar to Fig.~\ref{fig:clstars}, with the DM effects (both smooth and in halos) now added; we fix the DM channel to the muon one, the DM mass to 1 GeV, and $p_\mathrm{ann}=2.3\times10^{-7}$, saturating current bounds. While in Fig.~\ref{fig:cl_halos} only a single, realistic value $10^6$ for the boost factor is shown,  in Fig.~\ref{fig:cl_halos_res}, we present the residuals with respect to the reference $\Lambda$CDM models, with growing effect of the halo term (besides the 10$^6$ one, also a factor 10 and 100 larger). 
It is clear that, at least for the phenomenological model for the astrophysical reionization, potentially detectable effects emerge. 
Since we fixed $\theta_s$ and $\tau_\mathrm{reio}$, we nearly eliminated any oscillatory patterns and step-like discrepancy between the set of curves describing the TT spectrum.
This is not entirely true in models with a significant effect of annihilation in halos, for which a residual step-like effect can be clearly observed in Figure~\ref{fig:cl_halos_res}. This has to do with the ambiguity in defining $\tau_\mathrm{reio}$ in presence of partial reionisation at high-redshift, as explained in Appendix~\ref{sec:AppendixD}. By just fixing $\tau_\mathrm{reio}$, we do not eliminate completely the dependance of the high-$l$ temperature spectrum on the reionisation history. Apart from this effect, which could be compensated by a shift in the reionisation time or in the overall normalisation of the primordial spectrum, changes remain very modest and limited to small scales
 (where degeneracies with other parametrical extensions of $\Lambda$CDM are likely present). On the other hand, they can become sizable and characterized by a peculiar pattern at low-$l$ on the polarization power spectrum, coming from the enhanced rescattering of photons at intermediate  redshifts $z_\mathrm{reio} \leq z \leq 600$. While for realistic values of the halo parameters and DM mass it is still likely that this falls within the cosmic variance,  it cannot be excluded that more extreme values of the parameters could be independently ruled out by CMB polarization observables, provided that forthcoming final release of Planck  data manages to keep under control the systematics at low$-l$.

\section{Discussion and prospects}
\label{sec:disc}

The CMB temperature and polarization anisotropy pattern is sensitive to the energy injection by DM annihilation, essentially only 
via its constraints on the reionization optical depth $\tau_{\rm reio}$ (we clarified in appendix \ref{sec:AppendixD} in which limit this is actually true).
In turn, the optical depth $\tau_{\rm reio}$ probed by the CMB depends on the reionization history, i.e. the function $x_e(z)$, which itself couples to the thermal history of
the gas, controlled by the function $\TM(z)$ (their standard evolution equations have been reported in Appendix \ref{sec:appA} for completeness).
One major element of novelty of this work has been to study for the first time the dependence of the cosmological observables 
$x_e(z)$ and $\TM(z)$ from the underlying astrophysical reionization/heating source model, both in presence and absence of DM sources. We 
modified the dynamical equations for $x_e$ and $\TM$ (see Sec.~\ref{subsec:astro}) arguing that virtually all previous treatments have been 
incomplete and inconsistent in that respect.

We have then revisited  the problem of  current CMB constraints on dark matter annihilation in halos, with a state-of-the-art treatment. 
We have followed the by now standard formalism of Ref.~\cite{Slatyer09} and made use of the numerical tools provided in \cite{Slatyer12}, to compute precisely how the energy is deposited in the medium.
We have clarified and corrected a few mistakes in previous work and improved over them. In appendix \ref{app:fz}, we provide a detailed comparison of the many formalisms used in the literature to study DM annihilation in halos and carefully explain where are the few mistakes or approximations over which we have improved.  Appendix \ref{sec:appC} describes in details the parametric model of DM halo formation adopted in our calculations, although a generalization to different parameterizations is straightforward and would not alter
our conclusions.

In agreement with most previous literature, we have confirmed that with conventional assumptions on DM models and halo formation, DM annihilation fails to play a dominant role in reionizing the universe, and can at most coexist with reionization from stars. Only very unrealistic halos could give a significant contribution to the reionization optical depth, and even then, such models are hardly compatible with astrophysical measurements of the ionization fraction at low redshift.
No plausible DM model can produce a too high reionization optical depth, hence CMB measurements of $\tau_{\rm reio}$ do not provide additional constraints on the annihilation cross section. At face value, this means that previous constrains derived assuming DM annihilation in the smooth background only are robust and independent of uncertainties on structure formation. Note that the recent update of the tools to compute the energy deposition in the medium provided by~\cite{Slatyer15-2}, following the results of Ref. \cite{Slatyer13}
would not alter our conclusions: since the account for the new channel of energy loss through very low energy photons (with energy $<13.6$ eV) produced during the development of the electromagnetic cascade,  such a refinement would only make the impact of dark matter annihilations in halos even slightly weaker. We have checked this conclusion following the so-called "approximate" method described in Refs. \cite{Slatyer13, Slatyer15-2} which consists in withdrawing some power to the transfer functions of \cite{Slatyer12}, using a specific new table supplied in Ref. \cite{Slatyer15-2}. This method is, according to the authors, as precise as the new transfers functions and is better suited for our treatment.

To assess if these rather pessimistic conclusions are linked to the intrinsic insensitivity of the CMB to these effects or merely to the current lack of precision, in section~\ref{sec:cmb} we have computed the CMB angular power spectra, and compared the results with or without annihilation in halos.  We have shown that, {\it within standard assumptions for the astrophysical reionization} and for plausible halo models,  both the effects on the TT and EE multipole spectra are unobservable, falling below the level of cosmic variance.  One would conclude
that fits of CMB spectra in presence of DM annihilation in halos cannot provide (even in principle) better constraints on DM models. However, we have also shown that to some extent this conclusion can be altered if  a different scenario for the astrophysical source of reionization is adopted. At present, it cannot be excluded that some extreme but viable DM halo parameter
space might be eventually probed by CMB polarization data, provided errors can be kept at the level of the cosmic variance. To the best of our knowledge,
it is the first time that this effect is highlighted. 

\begin{figure}[!h]
\centering 
\includegraphics[scale=0.30]{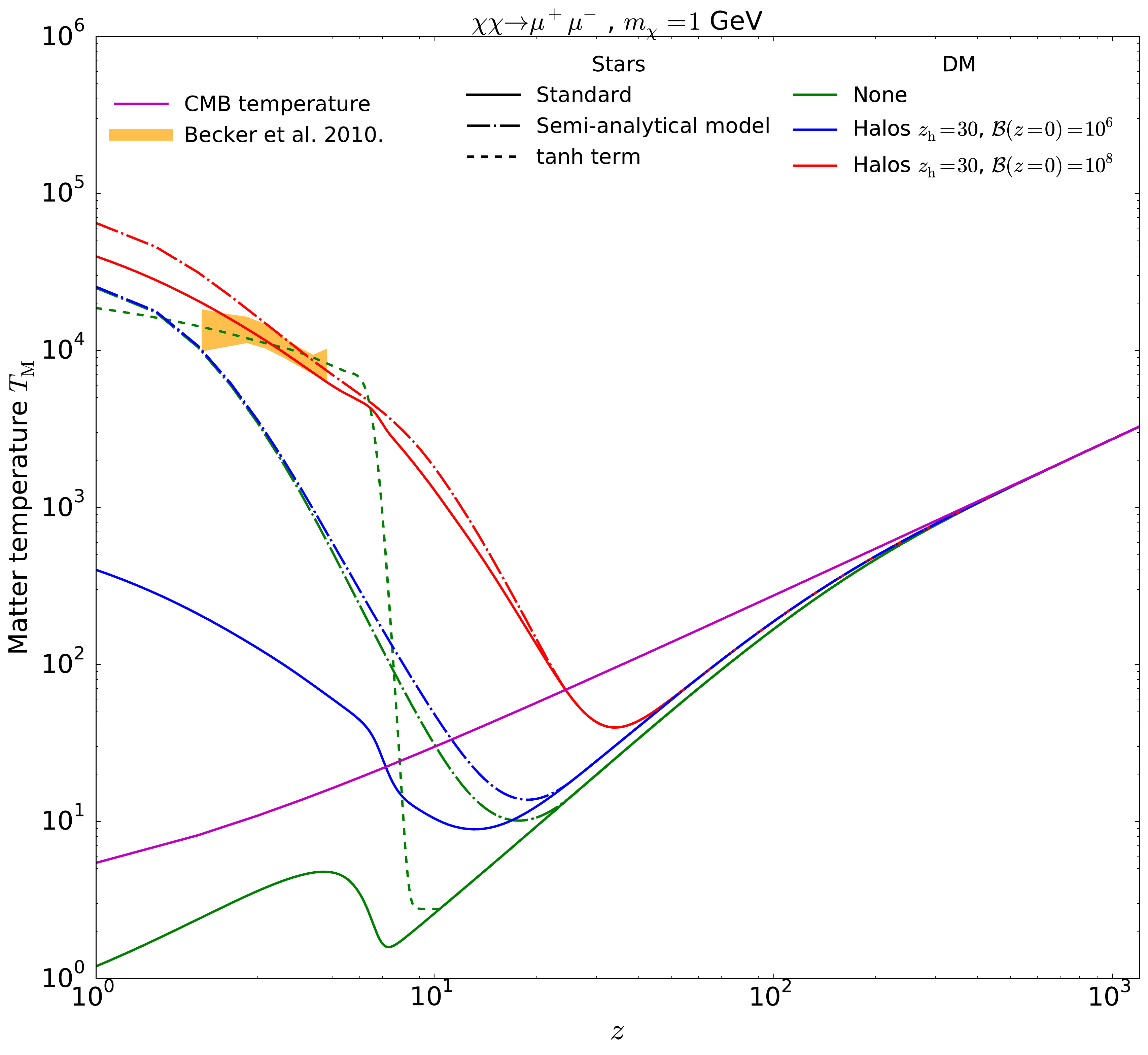}
\caption{\label{fig:TM_halo_mix} As in Fig.~\ref{fig:xe_halo_mix} (see also the right panel in Fig.~\ref{fig:StarsxeTM}) but for the IGM temperature.}
\end{figure}
Looking beyond the CMB probe, our results also leave the door open to some encouraging perspectives to further constrain DM annihilation in halos via other obserbables,
more directly linked to $x_e(z)$  or $\TM(z)$. To illustrate this point,  one can look for instance at Fig.~\ref{fig:TM_halo_mix},
where we report the evolution of $\TM$  for several mixed reionization models, analogous to what shown in Fig~\ref{fig:xe_halo_mix} for the evolution of $x_e(z)$. In particular,
green lines are benchmark evolutions for purely astrophysical scenarios (dashed: single step reionization/reheating; dot-dashed, phenomenological one; solid line: standard case where only the feedback from $x_e(z)$ is included in the dynamics of $\TM$). The blue and red versions of the corresponding lines show the case where both smooth DM injection and halo one have been added as well, with growing role of halos: DM annihilation in halos with $z_{\rm h} = 20,\,{\cal B}(z=0)=10^6$ are reported in blue, and $z_{\rm h} = 30,\,{\cal B}(z=0)=10^8$ in red, respectively.

The dot-dashed red line shows that already with current rudimentary constraints, if one could trust the phenomenologically motivated model for the astrophysical heating, extreme halo parameters could be excluded. At the same time, as shown by the solid green curve, it is immediately obvious that current treatments of the $\TM(z)$ evolution are strongly inadequate, and a significant effort should be put in achieving a realistic modeling of the sources of heating.
The qualitatively most interesting effect of DM annihilation in halos is the possibility of a sign shift in the difference between $T_\gamma$ and $\TM$ at much higher redshift than expected
for astrophysical models. This could be detectable for instance with 21cm surveys, since the signal would appear either in absorption or in emission (e.g.~\cite{Araya:2013dwa} and reference therein). Tomographic surveys of the cosmological 21cm observable, sensitive both to $x_e(z)$ and $\TM(z)$, are certainly the most promising avenue to progress in the knowledge of these redshift epochs.
Numerous experiments such as PAPER 64~\footnote{http://eor.berkeley.edu}, 21CMA~\footnote{http://21cma.bao.ac.cn}, MWA~\footnote{http://www.mwatelescope.org}, LOFAR~\footnote{http://www.lofar.org}, HERA~\footnote{http://reionization.org} or SKA~\footnote{http://www.skatelescope.org}, are now  (or will be) attempting at measuring the 21 cm signal.
Hopefully, it could be possible to see the impact of halos through such a sign shift at relatively high redshift, when the stars are not yet ``polluting'' the signal.
If not, a good description of the influence of stars on $\TM$ and $x_e$ would be of first importance, as we have illustrated using semi-empirical models taken from \cite{Bouwens15,Pober15,Robertson11,Robertson15}. Some authors are also trying to improve on the modelling of these effects using numerical simulation (such as {\sf 21cmFAST} \cite{21CMFAST,Valdes12,Evoli14}).
Achieving a sufficiently accurate treatment of the interplay of stellar and exotic sources is definitely a complicated task and a long term goal, but certainly it deserves further investigations.

\acknowledgments
We thank Marco Cirelli, Sergio Palomares Ruis and Aaron Vincent for very useful discussions. We also thank the anonymous referee for very constructive comments. Support by the Labex grant ENIGMASS is acknowledged.
\appendix
\section{Appendix A: Peebles's recombination \label{sec:appA}}

We want here to briefly introduce the standard tools useful for computing  observables relevant to the recombination and reionization epoch. 
The fundamental equations have been introduced by Peebles, quickly followed by Zel'dovich, Kurt and Sunyaev in the seminal papers \cite{Peebles,Zeldovich}. 
Since then, there has been substantial improvements of the model, as detailed in~\cite{Seager:1999bc,Mukhanov:2005sc,Weinberg:2008zzc,AliHaimoud:2010dx} and references therein. In the following, we will briefly sketch the basic ideas in order to keep our article self-contained. \\
In the standard picture, neutral hydrogen is not directly formed by capturing an electron in the ground state (1$s$), because this process yields a photon of the exact energy necessary to reionize a neutral atom, and hence results in no net decrease of $x_e$.
For a similar reason the 2$p$ level gives no significant net capture, since the emitted photons will re-excite another atom, that can then easily escape if the atom interacts with a CMB photon.
Recombination hence typically requires an electronic capture in the 2$s$ excited state, with a de-excitation through a {\em forbidden} transition, in order for these photons not to re-ionize other atoms.  
The accurate calculation of this process requires to follow coupled kinetic equations for the free electrons and thermal photons (taking into account a certain number of hydrogen levels), a problem typically solved numerically.
The most recent papers about recombination \cite{Rao:2015xpa,Desjacques:2015yfa}  distinguish between HII, HeIII and HeII recombinations. 
They have highlighted very interesting effects on the CMB spectrum itself, through deviations from the black-body shape known as the ``recombination lines'', which are in principle detectable with a dedicated experiment, but such refinements are not necessary for our purposes.

Assuming that helium has entirely recombined, which is the case in the redshift range we are interested in, the evolution equation for the free electron fraction $x_e \equiv n(\textrm{HII})/n(\textrm{H}) \equiv n(\textrm{HeII})/n(\textrm{He})$ in term of the redshift $z$ writes:
\begin{equation}\label{eq:x_e_bis}
\frac{dx_{e}(z)}{dz}=\frac{1}{(1+z)H(z)}(R(z)-I(z))\,,
\end{equation}
with
\begin{equation}\label{eq:RandI}
R(z) = C\bigg[\alpha_{H} x_e^2 n_H\bigg], \qquad I(z)  = C\bigg[ \beta_{H}(1-x_e)e^{-\frac{h\nu_\alpha}{k_b\TM}}\bigg]
\end{equation}
These two terms are respectively the standard recombination and reionization rates.
The first term encodes the probability that one free electron encounters an ionized hydrogen, is captured but \textit{not} directly in the ground state and finally decays from the $n = 2$ state to the ground state before being ionized.
The second term encodes the probability that a CMB photon redshifts at the Lyman-$\alpha$ transition frequency, and hits a neutral hydrogen in the $n=2$ state.
This standard scenario is known as Peebles "Case B" recombination. 
In this framework, the coefficient $C$ is given by:
\begin{equation}
C = \frac{1+K_{H}\Lambda_{H}n_{H}(1-x_e)}{1+K_{H}(\Lambda_{H}+\beta_{H})n_{H}(1-x_e)}
\end{equation}
where $\Lambda_{H} = 8.22458\,$s$^{-1}$ is the decay rate of the metastable 2$s$ level, $n_H(1-x_e)$ is the number of $H$ atoms in the ground state, and $K_{H} =\frac{\lambda^{3}_{\textrm{Ly}_{\alpha}}}{8\pi H(z)}$ accounts for the 
cosmological redshifting of Lyman-$\alpha$ photons.
Finally,  $\alpha_H$ and $\beta_H$ are the effective recombination and photoionization rates for principal quantum numbers $\geq$ 2 in Case B recombination (per atom in the 2$s$ state), $\nu_\alpha$ is the Lyman-$\alpha$ frequency and $\TM$ is the temperature of the baryonic gas.\\
The evolution of the matter temperature is instead ruled by the equation:
\begin{equation}\label{eq:TM_bis}
\frac{d\TM}{dz}  =  \frac{1}{1+z}\bigg[2\TM+\gamma(\TM-\TCMB)\bigg]
\end{equation}
where the dimensionless parameter $\gamma$, also called opacity of the gas, is defined as :
\begin{equation}
\gamma \equiv \frac{8\sigma_T a_r \TCMB^4}{3Hm_e c}\frac{x_e}{1+f_{He}+x_e}
\end{equation}
with $\sigma_T$ the Thomson cross-section, $a_r$ the radiation constant, $m_e$ the electron mass, $c$ the speed of light, and $f_{He}(\equiv Y_p/[4(1-Y_p)])$ the fraction of helium by number of nuclei.
The first term is an adiabatic cooling due to the Universe expansion (for a massive non-relativistic particle the kinetic energy $E_k = p^2/2m \propto a^{-2}$ and can be related to the averaged energy of indivual particles in a monoatomic gas $E_k = 3/2 k_b\,T$), whereas the second one accounts for interactions between CMB photons and matter, mainly through Compton scattering with free electrons.
The evolution of $\TM$ therefore mainly depends on the second term, with the sign of the difference $\TCMB-\TM$ determining if collisions heat or cool down the gas, and the value of $\gamma$ responsible for the coupling strength:
If $\gamma \gg 1$, the second term dominates and Compton scatterings thermally couple CMB photons and baryons ($\TM \propto \TCMB \propto (1+z)$), whereas $\TM$ adiabatically decays as $\TM \propto (1+z)^2$ when $\gamma \ll 1$.
\section{Appendix B: Remarks on the efficiency function $f(z)$ \label{app:fz}}

The computation of the $f(z)$ for each DM model (mass and final state) is the key ingredient needed to correctly describe impact of DM annihilation on the cosmic gas. 
For typical WIMP models, this computation was made in \cite{Slatyer09, Slatyer12, Slatyer13}, and here we will only sketch the procedure. 
We refer reader to previous references for detailed explanations and estimates of the systematic uncertainties relative to the treatment. \\
In order to compute these functions, one needs to follow the evolution of an electromagnetic cascade starting at the DM mass (typically $\mathcal{O}$(GeV-TeV)) until the energy at which atomic processes dominate (sub-keV) and over the entire redshift range that we are interested in, typically from $z\sim$ few thousands until today.
This makes $\gtrsim 10$ orders of magnitudes in energy and 3 orders of magnitude in redshift that the code should be able to span.

A significant simplification can be achieved by splitting the cascade in two regimes: at high energy,  interactions with CMB photons dominate and are responsible for
particle multiplication and energy degradation; at low energies, electromagnetic particles interact with the gas and lead to  heating, excitation and ionization of the medium.
In practice, at each redshift, the first part of the computation provides the fraction of energy available to interact with the matter (which, going beyond the on-the-spot approximation, can be bigger than 1), namely the efficiency factor $f(z)$, whereas the second part of the computation yields the energy repartition fractions $\chi_j$. \\

The last remark we want to make consists in a comparison of the formalism used by authors to implement the energy deposition from DM annihilations.
Instead of parametrizing the energy deposition history through the $f(z)$ functions, the authors of \cite{Hooper09,Cirelli09,Natarajan10,Giesen} have used a more explicit method, which has however some caveats. 
Following notations of \cite{Cirelli09}, equivalent to other articles, one can write after a few manipulations the energy deposition as:
\begin{eqnarray}\label{eq:cirelli}
&&\frac{dE}{dVdt}\bigg |_{\textrm{dep}}(z) =\\
&&\int_z^\infty\frac{dz'}{(1+z')H(z')}\frac{(1+z)^3}{(1+z')^3}\frac{n_A(z)\sigmav}{2\mDM^2}\rho^2(z')\int^{\mDM}_{0} dE_\gamma E_\gamma\frac{dN_\gamma}{dE_\gamma'}(E_\gamma')e^{-\kappa(z',z;E_\gamma')}\sigma(E_\gamma) \nonumber
\end{eqnarray}
where $\sigma(E_\gamma')$ is obtained by summing $\sigma_{\gamma A\to e^-A+}$, $\sigma_{\gamma e^-\to \gamma e^-}$ and $\sigma_{\gamma A\to e^\pm A}$ (the index $A$ refers to H and He atoms), $n_A(z)$ scales as $(1+z)^3$, $\frac{dN_\gamma}{dE_\gamma'}(E_\gamma')$ is the spectrum of photons produced by the DM annihilations (both prompt and from inverse Compton scattering), $E_\gamma'$ is the energy at the time of injection and  $E_\gamma$ is the energy at the time of deposition,
 and $\kappa(z',z;E_\gamma')$ plays the role of an absorption coefficient defined as:
\begin{equation}
\kappa(z,z',E_\gamma) \simeq \int_{z'}^{z}dz''\frac{dt}{dz''}n_A(z'')\sigma(E_\gamma'')\quad .
\end{equation}
Three comments can be made about this formula.
\begin{itemize}
\item
First of all, a mistake is made in eq.~(4) of Ref.~\cite{Natarajan10} (the equivalent of our eq.~(\ref{eq:cirelli})) and that mistake propagated to Ref.~\cite{Giesen}. Indeed, in eq.~(4) of Ref.~\cite{Natarajan10}, the last integral is performed over $E_\gamma'$ (the energy at the time of injection) instead of $E_\gamma$ (the energy at the time of deposition). Since the two are related by $E_\gamma'=E_\gamma(1+z')/(1+z)$, the final result differs by a factor $(1+z)^2/(1+z')^2$ in the last integral over $dz'$. This can be seen explicitely in eq.~(4.21) of Ref.~\cite{Giesen}, where the exponent 6 should in fact be 8.
\item
Secondly, in order to perform the integral analytically, Refs.~\cite{Natarajan10,Giesen} assume that it is possible to consider that all interactions between injected photons and matter are due to Compton scattering in the Thomson limit. This approximation works well for the smooth background, but it is inaccurate compared to the approach of Slatyer et al.~\cite{Slatyer09}  in the context of annihilation in halos.

\item
To further compare the approach of Refs.~\cite{Hooper09,Cirelli09,Natarajan10,Giesen} with that of Ref.~\cite{Slatyer09}, it is possible to manipulate eq.~(\ref{eq:cirelli}) in such way to absorb all the energy and redshift dependance. Then, the equivalent of the energy deposition function $f(z)$ appears, defined as:
\begin{equation} \label{eq:fzcirelli}
\frac{dE}{dVdt}\bigg |_{\textrm{dep}}(z)  = \tilde f(z)   \rho^2(z)\frac{\sigmav}{\mDM}\,,
\end{equation}
where
\begin{equation} \label{eq:cirelli2}
\tilde f(z)=\int_z^\infty\frac{dz'(1+z)^5}{H(z')(1+z')^6}\frac{n_A(z) \rho^2(z')}{2\,\mDM\rho^2(z)}\int^{\mDM}_{0} dE_\gamma' E_\gamma' \frac{dN_\gamma}{dE_\gamma'}(E_\gamma')e^{-\kappa(z',z;E_\gamma')}\sigma(E_\gamma)\,.
\end{equation}
Even if this formalism goes one step beyond the ``on-the-spot'' approximation thanks to the integral over $z'$, the complexity of the cascade evolution is not taken into account. 
The two approaches do not give significantly different results when considering only the smooth background, since the CMB experiments are not sensitive to the shape of $f(z)$ at low $z$, but using eq.~(\ref{eq:cirelli2}) can lead to wrong results when halos are turned on.
\end{itemize}

\section{Appendix C: The boost function ${\cal B}(z)$ \label{sec:appC}}

Generally, it is possible to re-write the energy deposition rate in halos by introducing an enhanced dark matter density $\rho^2(z)=  (1+\mathcal{B}(z))\bar{{\rho}}^2(z)$. One now has:
\begin{equation}
\frac{dE}{dVdt}\bigg|_{\textrm{inj, halos}}=\rho^2_cc^2\Omega^2_{\textrm{DM}}(1+z)^6\frac{\sigmav}{\mDM}(1+\mathcal{B}(z))
\end{equation}
To compute this "boost" of the density, several ways have been introduced. 
In the so-called "Halo model" (HM) framework,  it is assumed that all the mass in the Universe is contained in virialized objects (the halos), fully characterized by their mass.
The key ingredients are the spatial distribution of matter inside a halo (usually called the \textit{density profile}) and the number of halos per unit mass (namely, the \textit{mass function}, that can evolve with the redshift).
This allows one to express the boost $\mathcal{B}(z)$ as  :
\begin{equation}
\mathcal{B}(z) = \frac{1}{\rho^2_c\Omega_{\textrm{DM}}}(1+z)^3\int_{M_{\textrm{min}}}^{\infty}dM\frac{dn}{dM}(z)\bigg(\int_0^{r_{200}}dr4\pi r^2\rho^2_{\textrm{h}}(r)\bigg)
\end{equation}
where $M_{\textrm{min}}$ is the minimal mass of DM halos and $r_{200}$ the radius of a sphere enclosing a mean density equal to 200 times the background density (some authors prefer to use the virial radius $r_{\rm v}$, but they are strictly equivalent).
The last integral is usually recast in terms of the concentration function $F(\ch)$, which depends on the concentration parameter $\ch \equiv r_{200}/r_s$ with $r_s$ the scale radius of the given profile: 
\begin{equation}
\int_0^{r_{200}}dr4\pi r^2\rho^2_{\textrm{h}}(r) = \frac{M\bar{\rho}(z_F)}{3}\bigg(\frac{\OmegaDM}{\OmegaM}\bigg)^2F(\ch)
\end{equation}
where $z_F$ is the redshift of halo formation (not yet well known) and $\bar{\rho}(z_F) = 200\rho_c\OmegaM(1+z_F)^3$ the average matter density within a sphere of radius $r_{200}$.\\
One now needs to relate the concentration function to the halo profile, and several types of profile have been proposed in the litterature.
For instance, the most commonly used are the Navarro-Frenk-White (NFW) profile \cite{Navarro:1995iw}, Einasto profile \cite{Navarro:2008kc,Graham:2005xx} and isothermal-like Burkert profile \cite{Burkert:1995yz}, the former two showing a more peaked  distribution than the latter.
The last key quantity -- the mass function --  which in this framework also encodes how the density evolves with the redshift, can be calculated analytically in the Press-Schechter formalism \cite{Press:1973iz}, leading to :
\begin{equation}
\frac{dn}{dM}=\frac{\rho_c\OmegaM}{M}\frac{d\ln\sigma^{-1}}{dM}f(\sigma)
\end{equation}
where the variance of the density field $\sigma$ and the differential mass function $f(\sigma)$ have been introduced.
We shall not develop further the HM and the Press-Schechter formalism and refer to \cite{Giesen} for more details. 
Following their development, one can show that, in the frame of the HM with the Press-Schechter prescription for the mass function, the boost from halos is given by:
\begin{equation}\label{eq:BoostFactor}
\mathcal{B}(z)=\frac{f_{\text{h}}}{(1+z)^3}\text{erfc}\bigg(\frac{1+z}{1+z_\text{h}}\bigg)
\end{equation}
where $f_{\text{h}}=\frac{200}{3}(1+z_F)^3F(c_{\text{h}})$ and $z_\text{h}$ is the characteristic redshift at which halos start to contribute (we typically have $z_F\simeq2z_\text{h}$). 
It could be possible to use more advanced prescription for the halo mass function, such as the Sheth-Tormen formula \cite{Sheth:1999mn} or the one from Tinker et al. \cite{Tinker:2008ff}, but this would not lead to major differences for the problem we are dealing with. 
Indeed, as it appears in this framework, the mass function mainly dictates the evolution of the distribution with $z$ and is only known for very small redshifts compared to the range we want to span. 
The error function is merely a  prescription inspired by a simple model to describe the transition from a smooth distribution of matter to structures of virialized objects, whose normalization at low redshift is  encoded in the function $f_{\text{h}}$. 
To learn how structure formation precisely happens, especially at high $z$, is one ingredient that would allow one to improve over the present study.
Here, we will limit ourselves to vary $z_{\textrm{h}}$ in the interval [20,30] to ease the comparison with existing literature. Concerning the additional parameter  $f_{\text{h}}$, we deduce a reasonable
range for it from the range of $\mathcal{B}(z=0)$ computed in~\cite{Serpico:2011in,Sefusatti:2014vha}. In those references, an alternative method is used
to compute the boost factor, by direct integration of the power-spectrum down to very small scales (where a cutoff is induced by WIMP dark matter
free-streaming or kinetic decoupling) calibrated directly on simulations. The large uncertainties
in dealing with the power-spectrum in the deeply non-linear range explain the broad interval $\mathcal{B}(z=0)\in[10^4,10^8]$
inferred for typical WIMP candidates, although more recent simulations tend to narrow it down to $[5\times 10^4,10^6]$.

\section{Appendix D: Discussion on $\tau_{\rm reio}$ as it is measured by Planck}
\label{sec:AppendixD}

The Planck collaboration explains in sec 3.4 of Ref.~\cite{Planck15} how they proceed to obtain a measurement of the optical depth to reionization $\tau_{\rm reio}$.
Let us recall briefly how this is done, and why this might be problematic when one modifies the reionization history in a non-trivial way and wants to compare the new $\tau_{\rm reio}$ to the Planck results.

In general, the optical depth at redhift $z$ is defined as 
\begin{equation}
\tau(z) \equiv \int^{z}_{0} n_H(z) x_e(z) \sigma_T\frac{dt}{dz'} dz'~.
\end{equation} 

Planck obtains an estimate of the reionization optical depth by: (i) assuming single-step reionization where $x_e(z)$ is described by a postulated function centered on an adjustable redshift $z_\mathrm{reio}$; (ii) computing the corresponding CMB temperature and polarization power spectra using the Bolztmann codes CAMB~\cite{camb} (or CLASS~\cite{Lesgourgues:2011re,Blas:2011rf} for cross-checks); (iii) comparing these spectra with those extracted from observed maps. The fit directly gives some bounds on the reionization redshift $z_\mathrm{reio}$. At the same time, for each model, the codes also compute the integrated parameter $\tau_{\rm reio}^\mathrm{camb} \equiv \tau(z_{\rm c})$,
where the cut-off redshift $z_{\rm c}$ takes an arbitrary default value $z_{\rm c} = 40$ in both CAMB and CLASS, and reports bounds on this $\tau_{\rm reio}^\mathrm{camb}$. Actually, since there is a one-to-one correspondance between $\tau_{\rm reio}^\mathrm{camb}$ and $z_\mathrm{reio}$ (valid for the postulated category of reionization models), the Bayesian parameter extraction can even be done with a flat prior on $\tau_{\rm reio}^\mathrm{camb}$ rather than $z_{\rm reio}$.

In such models, reionization affects:
\begin{itemize}
\item the low-$l$ part of the polarization spectra, creating the so-call reionization bump, 
\item the high-$l$ part of the temperature and polarization spectra, which are step-like suppressed, with a suppression factor saturating above $l \sim 70$ at a value very close to $e^{-2\tau_{\rm reio}^{\rm camb}}$.
\end{itemize}
The first impact is very challenging to detect because of the smallness of the signal, which requires a very good control of instrumental systematics and polarized foreground emissions. Even if theoretically, the signature of reionization on the CMB polarization spectra is very clear, the actual measurement of the $E$-mode low-$l$ spectrum still has quite large uncertainties, and that of the $B$-mode spectrum is not sensitive enough to probe any reionization bump. Hence, current sensitivity to reionization comes mainly from the second of the two effects. The data probe mainly the integrated parameter $\tau_{\rm reio}^\mathrm{camb} $, and the bounds reported on this parameter are often thought to be model-indepent (i.e., valid for any reionization history).
 
\begin{figure}[!h]
\centering
\includegraphics[scale=0.33]{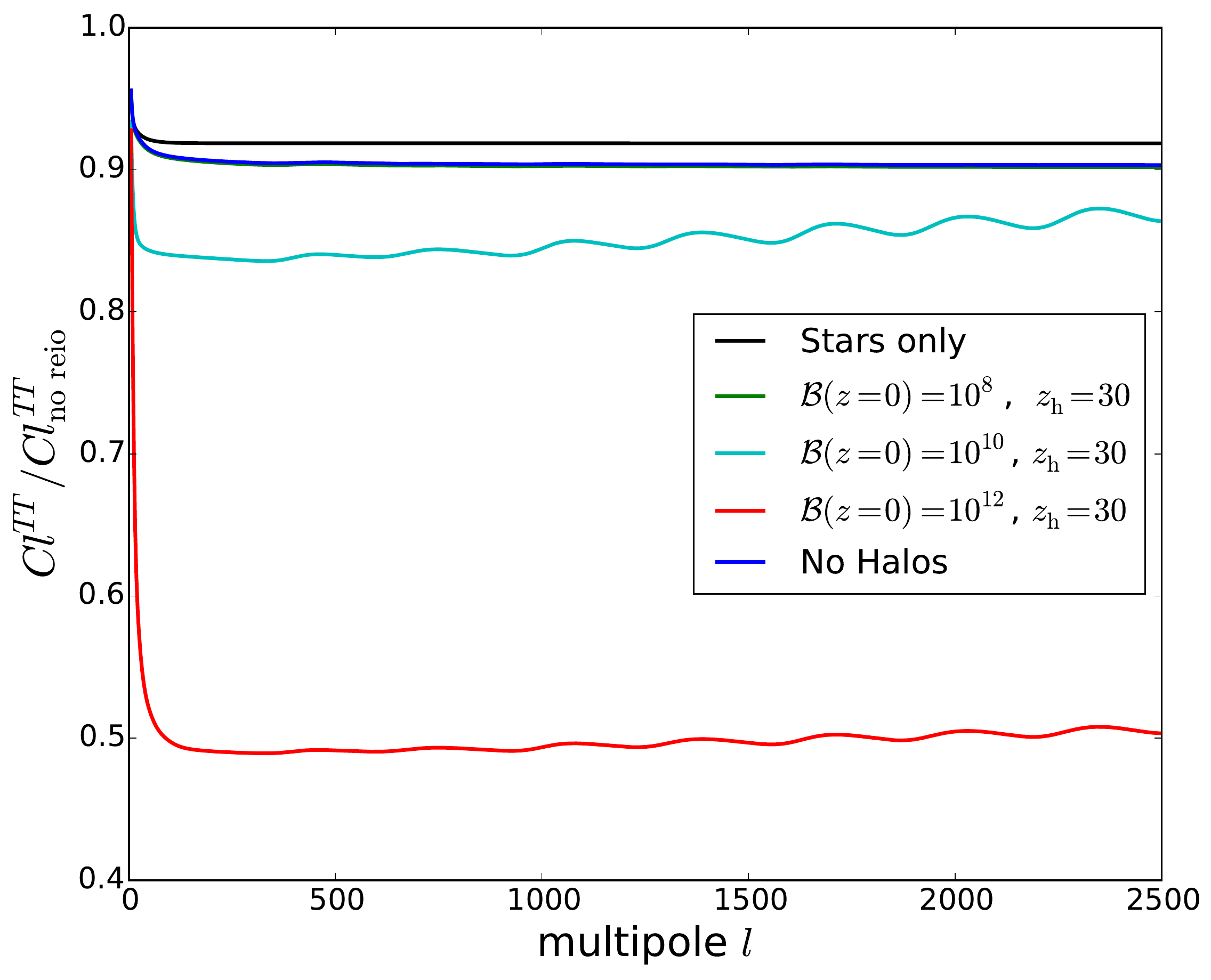}
\includegraphics[scale=0.33]{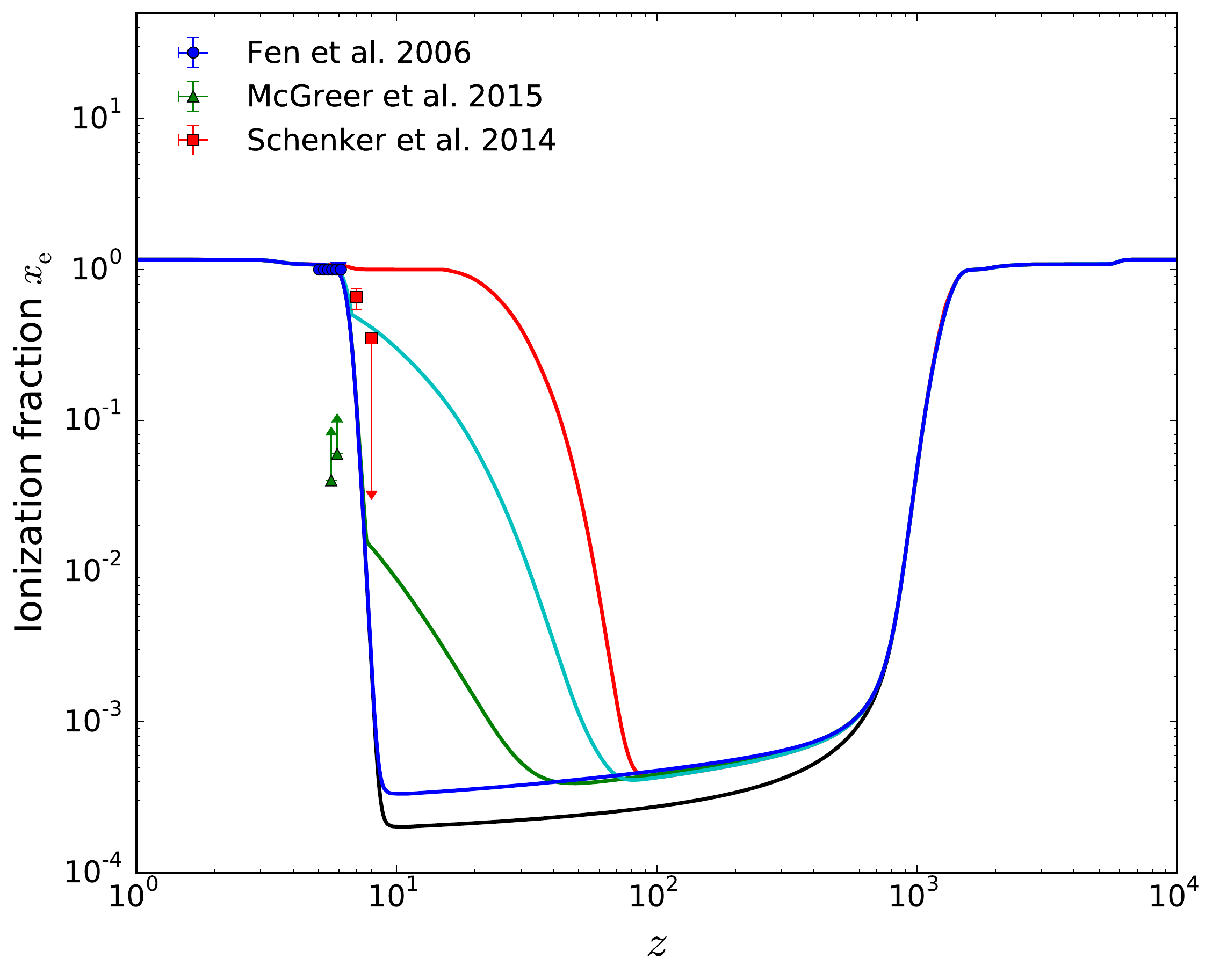}
\caption{Left pannel - Comparison of the CMB temperature power spectrum for several mixed star-DM reionization models, with $z_{\rm reio} = 6.5$ and a DM particle of 1 GeV annihilating into electrons, normalized to the CMB power spectrum in a universe without reionization. Right pannel - The corresponding ionization histories.\label{fig:ComparaisonCl}}
\end{figure}

However, some of the above statements are only true for models such that the ionization fraction $x_e(z)$ only starts raising at low redshift. For more complicated models like those involving DM annihilation, the fact that there is a step-like suppression of the $C_l$'s, and that the suppression factor asymptotes to $e^{-2\tau_{\rm reio}^{\rm camb}}$, are not necessarily accurate. For instance, Figure \ref{fig:ComparaisonCl} shows that in presence of DM annihilation in halos, the reionization effect is not exactly step-like. It has superimposed oscillations, and the suppression factor drifts significantly between $l \sim 100$ and $l \sim 2500$.

\begin{figure}[!h]
\centering
\includegraphics[scale=0.45]{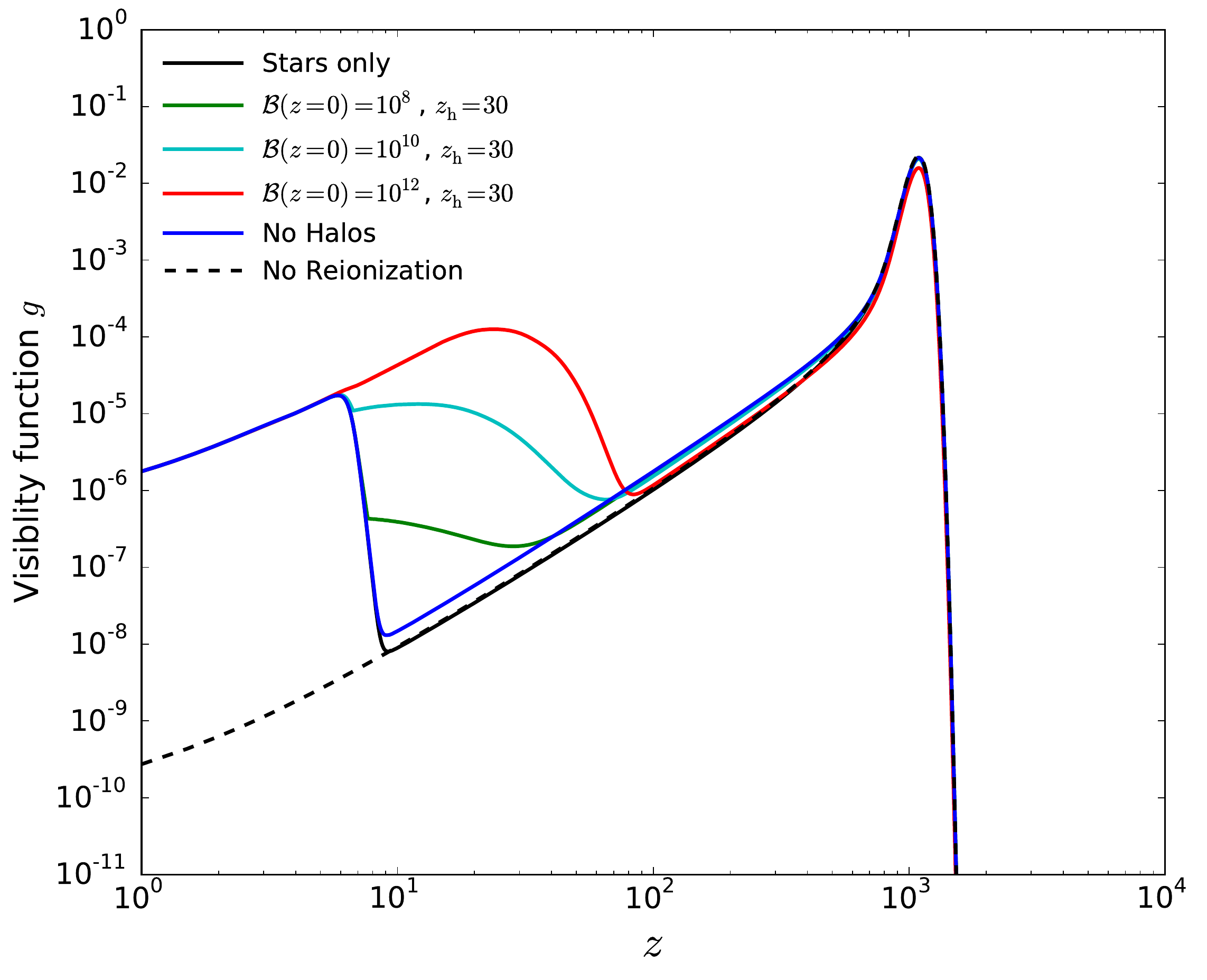}
\caption{Visibility function for the reionization history from fig.~\ref{fig:ComparaisonCl}, compared to the visibility function in the absence of reionization.\label{fig:Visibility}}
\end{figure}

This can easily be understood using the line-of-sight integral formalism~\cite{Seljak:1996is}, which shows that the Sachs-Wolfe and Dopper terms in the CMB transfer functions are given by an integral over time of a product of cosmological perturbations, Bessel functions, and finally the visibility function $g \equiv \tau' e^{-\tau}$ (as well as its  time derivative in the Doppler term). The visibility function $g(z)$ gives the probability of last scattering, and is normalised to one by construction: $\int g(z) \, dz =1$. In absence of reionization, $g(z)$ has a single peak at recombination, and it could be approximated by zero for $z \ll z_\mathrm{dec}$ without changing the result (physically, this means that all observed photons last scattered near $z=z_\mathrm{dec}$). With conventional single-step reionization model (black line in fig.~(\ref{fig:Visibility})), a second peak centered near $z=z_\mathrm{reio}$ is visible. For high $l$'s, the integral over time still only picks up contributions near $z=z_\mathrm{dec}$. However, the normalization of $g(z)$ implies that the amplitude of the recombination peak in $g(z)$ is reduced by exactly $e^{-\tau(z)}$, where $z$ can be choosen anywhere in the range $z_\mathrm{reio} \ll z \ll z_\mathrm{dec}$ (e.g. $z=40$, as taken by default in CAMB and CLASS). Hence the high-$l$ temperature and polarization power spectra are supressed by exactly $e^{-2\tau_{\rm reio}^{\rm camb}}$.

This conclusion is almost unspoiled by the introduction of DM annihilation in the smooth background, or for relatively small halo contribution (blue and green curves in fig.~(\ref{fig:Visibility})).
However, the non-standard shape of $g$ in the case where reionization in DM halos is very strong and starts early (cyan and red curves in fig.~(\ref{fig:Visibility})) produces non-trivial effects, because the line-of-sight integral starts to pick up sizeable contributions coming from the second reionization bump. Physically, this comes from extra small-scale anisotropies generated by CMB photons rescattering at medium redshift, for instance between $z\sim 40$ and $z\sim 100$.  

Since Planck measures the optical depth mainly throught its suppression effects on the $C_l$'s at high $l$, it is possible to say that what Planck really measures is an effective parameter
$\tau_{\rm eff} \equiv - \frac{1}{2} \log (C_l^{TT} / C_l^{TT,~{\rm no~reio}})$, where $l$ is some effective multipole value in the region where the data has maximum sensitivity, and we are assuming that
most of the sensitivity comes from temperature. For usual single-step reionization models, $\tau_{\rm eff}$ coincides both with $\tau_{\rm reio}^\mathrm{camb}$, computed up to $z_{\rm c}=40$, and with a more sensible and robust
definition that one could give of the reionization optical depth, where the integral would run up to the redshift at which the ionization fraction is minimal: 
\begin{equation}
\tau_{\rm reio}^\mathrm{min} \equiv \tau(z_*)~, \quad \quad x_e'(z_*)=0~.
\end{equation}
For models with a raise in  $x_e(z)$ at high redshift, it is obvious that $\tau_{\rm eff}$ and $\tau_{\rm reio}^\mathrm{camb}$ can be very different, because integrating up to $z_{\rm c}=40$ is not sufficient. We also checked this numerically and found a significant mismatch. A more interesting question is whether $\tau_{\rm eff}$ and $\tau_{\rm reio}^\mathrm{min}$ are close to each other. In other words, can we still say that the total effect on the CMB spectra at high $l$ is related to the optical depth computed up to the very beginning of reionization? We tested this Ansatz explicitly, and found that the difference between these two quantities is small, being of the order of 10\% in the worst cases, i.e., still within the error bars of Planck. In summary:
\begin{itemize}
\item for usual single-step reionization: $$\tau_{\rm reio}^\mathrm{camb} = \tau_{\rm eff} = \tau_{\rm reio}^\mathrm{min}\,.$$
\item in more general cases with an early enhancement of the ionization fraction: $$\tau_{\rm reio}^\mathrm{camb} \neq \tau_{\rm eff} \simeq \tau_{\rm reio}^\mathrm{min}\,.$$
\end{itemize}
Therefore, as a first approximation, it is safe to define everywhere in our analysis $\tau_{\rm reio}$ as $\tau_{\rm reio}^\mathrm{min}$, and to compare it with the value of $\tau_\mathrm{reio}$  reported by Planck. This conclusion legitimates the analysis performed in section~ \ref{sec:reio}. For even better accuracy, the next step would be to perform a full parameter extraction in which exact power spectra are fitted to the data, and to derive new bounds on $\tau_\mathrm{reio}$ valid in all cases, but this is not necessary for reaching the main conclusions of this paper.

\bibliographystyle{ieeetr}

\bibliography{biblio}
\end{document}